\documentclass[]{aa}  

\usepackage{hyperref}
\usepackage{graphicx}
\usepackage[dvipsnames]{xcolor}
\usepackage{ulem}
\usepackage{txfonts}
\newcommand{\teff}{$T_{\rm eff}$}

\begin{document}

   \title{Matching seismic masses for RR Lyrae-type and oscillating red horizontal-branch stars in M4}



   \author{L\'aszl\'o Moln\'ar\inst{1,2}
          \and
          Henryka Netzel\inst{3,4}
          \and
          Madeline Howell\inst{5,6}
          \and
          Csilla Kalup\inst{1,2}
          \and
          Meridith Joyce\inst{1,7}
          }

   \institute{Konkoly Observatory, HUN-REN Research Centre for Astronomy and Earth Sciences, MTA Centre of Excellence, Konkoly-Thege Mikl\'os \'ut 15-17, H-1121, Budapest, Hungary\\
              \email{molnar.laszlo@csfk.org; lmolnar@konkoly.hu}
        \and
            ELTE E\"otv\"os Lor\'and University, Institute of Physics and Astronomy, 1117, P\'azm\'any P\'eter s\'et\'any 1/A, Budapest, Hungary
         \and
         Institute of Physics, \'Ecole Polytechnique F\'ed\'erale de Lausanne (EPFL), Observatoire de Sauverny, 1290 Versoix, Switzerland
         \and
            Nicolaus Copernicus Astronomical Centre, Polish Academy of Sciences, Bartycka 18, PL-00-716 Warszawa, Poland
        \and
            Department of Astronomy, The Ohio State University, 140 W. 18th Ave., Columbus, OH 43210, USA
        \and
            Center for Cosmology and Astroparticle Physics (CCAPP), The Ohio State University, 191 W. Woodruff Avenue, Columbus, OH 43210, USA
         \and
            University of Wyoming, 1000 E University Ave, Laramie, WY USA
         }

   \date{}

 
\abstract{Globular clusters offer a powerful way to test the properties of stellar populations and the late stages of
low-mass stellar evolution. 
In this paper we study oscillating giant stars and overtone RR Lyrae-type pulsators in the nearest globular cluster, M4, with the help of high-precision, continuous light curves collected by the \textit{Kepler} space telescope in the K2 mission. We determine the frequency composition of five RRc stars and model their physical parameters from linear pulsation models. We are able, for the first time, to compare seismic masses of RR Lyrae stars directly to the masses of the very similar red horizontal branch stars in the same stellar population, independently determined from asteroseismic scaling relations. We find average seismic masses of $0.648\pm0.028\,M_\odot$ for RR Lyrae stars and $0.657\pm0.034\,M_\odot$ for red horizontal-branch stars. While the accuracy of our RR Lyrae masses still relies on the accuracy of evolutionary mass differences of neighboring horizontal branch subgroups, this result strongly indicates that RRc stars may indeed exhibit high-degree, $l = 8$ and 9 non-radial modes, and modeling these modes can provide realistic mass estimates. We compare the seismic masses of our red horizontal branch and RR Lyrae stars to evolutionary models and to theoretical mass relations and highlight the limitations of these relations.  }


   \maketitle
%

\section{Introduction}
\label{sect:intro}
Masses of RR Lyrae stars are notoriously difficult to determine. This is a problem, because the masses of Horizontal Branch (HB) stars, and in particular the masses of RR Lyrae stars, located at the intersection of the HB and the instability strip of classical pulsating stars, are crucial pieces of information for stellar evolution theories and models \citep{Catelan-2009}. Masses would provide information not only on the mass distribution along the HB, but also on the amount of mass loss during the red giant phase of stellar evolution and on the accuracy of mass estimates based on pulsational envelope models, as well. Thus, many avenues of research have been already explored to determine the masses of RR Lyrae stars. 

Binarity, for example, would be a straightforward way to infer dynamical masses. However, not only do we not know any eclipsing binaries with an RR Lyrae member, it is very difficult to find systems that contain an RR Lyrae star at all \citep{wade-1999,kennedy2014,Hajdu-2015,Liska-2016}. Since \textit{bona fide} RR Lyrae stars are considered to be very old, their companions have either evolved into white dwarfs or other stellar remnants already, or must be smaller stars, most likely still evolving along the lower main sequence as K--M dwarfs. As such, the contribution of RR~Lyr companions to the total light output of the system is expected to be minimal, precluding the detection via spectral energy distribution, as double-lined spectroscopic binaries.

The dearth of eclipsing RR~Lyrae stars can also be explained by the evolutionary pathways of these pulsators. These stars moved past the red giant phase without any significant interactions with a companion that could have directed the primary away from evolving onto the HB and towards binary evolution scenarios. Companions thus must be distant to avoid any mass transfer, and a distant companion orbiting a now shrunken horizontal-branch star would make the probability of observing the expected shallow eclipses from the system very low. The only eclipsing binary candidate system turned out to be an impostor, a Binary Evolution Pulsator \citep{BEP-2012,BEP-2017}. In this case, mass transfer directed a much lighter primary to rapidly cross the instability strip and to experience RR Lyrae-like pulsations. We note that binarity and a limited amount of mass transfer has been proposed to explain metal-rich RR Lyrae stars, but those systems are expected to have wide orbits, too \citep{bobrick-2024,Abdollahi-2025}.

More indirect methods have shown increased promise, but all have their own caveats. For example, radial velocity searches are hindered by the expected long orbital periods and small signals relative to the much larger pulsation velocities, and TU UMa remains the only viable candidate \citep{Liska-TUUMa-2016}. A recent work proposed new binary candidates, but those still require follow-up and confirmation \citep{barnes-2021}. 

An alternative way to detect binaries and estimate masses is the light-time effect, but this method has been impeded by the prevalence of the Tseraskaya--Blazhko effect\footnote{Ms.~Tseraskaya (Ceraski) collected the data at Moscow Observatory that led to the discovery.}, the quasi-periodic modulation of the pulsation amplitude and phase \citep{blazko1907AN,Shapley-1916,kurtz2022}. Nevertheless, many Galactic stars show cyclic phase variations that can be caused by light-time effect \citep[see, e.g.,][]{Liska-2016,Sylla-2024}. Furthermore, several binary candidates have been identified in the OGLE survey this way \citep{Hajdu-2015,Hajdu-2021}. And while \citet{Hajdu-2021} did not address the masses of the primaries directly, these authors found a trimodal distribution in the mass functions of the secondaries. 

As we have shown, dynamical methods to determine masses of RR Lyrae stars suffer from various theoretical and/or practical shortcomings. This limits us to other, more indirect, model-dependent methods. Fundamental physical parameters such as luminosity, \teff{} and $\log g$, along with elemental abundances, can be fitted with evolutionary tracks and atmosphere models. A relation to calculate masses from other physical parameters (period, luminosity, \teff{}) was first proposed by \citet{vAB1971} and by multiple authors since. However, they come with strong uncertainties in various model parameters that are very loosely constrained by observations, if at all. 
These include the He content of the models, the approximations chosen for internal processes like convection and overshoot, or various aspects of mass loss along the evolutionary pathway \citep[see, e.g.,][]{marconi2018,anders-2023,Joyce-MLT-2023,Joyce-2024}. \citet{nemec-2011} used such relations to calculate masses for the RR Lyrae stars in the \textit{Kepler} field, for example, but without providing uncertainties for them. 

One way to limit uncertainties in stellar parameters is to study RR Lyrae stars in globular clusters. There are currently close to three thousand RR Lyrae stars known in 115 Galactic globular clusters \citep{Cruz2024}. For these stars, parameters like distance and metallicity can be determined accurately for the members, and using that information, the \citet{vAB1971} relation was employed in many cases to estimate masses. M3, for example, was studied by multiple authors, and \citet{Valcarce-2008} determined the mass distribution for the entire HB of the cluster based on evolutionary models, and found the distribution to be potentially bimodal. 
However, they discuss how that finding can be complicated by limitations present in the stellar tracks. A different approach was taken by \citet{kumar-2024}, who fitted the light curves of the M3 RR~Lyrae variables directly, based on a grid of non-linear pulsation models. Direct light curve fitting is a promising technique \citep[see also,][]{Bellinger-2020,Das-2025}, but current one-dimensional non-linear pulsation models still cannot reproduce light curves very accurately, due to limitations in handling convection, for example \citep{kovacsgb-2023}. 

Uncertainties in evolutionary and non-linear pulsation models can also be mitigated if we turn towards asteroseismology, and model the frequencies of the oscillatory modes observed in the star \citep[see, e.g.,][]{kjeldsen-1995,aerts-2021}. Incorporating asteroseismic constraints has shown great potential recently for stars outside the instability strip. However, classical pulsators like RR Lyrae stars face obstacles in this aspect, too: in these stars, most often only a single mode can be observed, severely limiting our ability to constrain physical parameters from frequencies alone. Masses for a few double-mode RR~Lyrae stars have been calculated from non-linear non-adiabatic hydrodynamical simulations, \citep{Molnar-rrd-2015}, but, as mentioned before, the accuracy of non-linear models have been called into question \citep{smolec-2008}. 

More recently, a new method based on the discovery of low-amplitude signals in overtone RR Lyrae stars has been proposed \citep{netzel-2015-061a,netzel-2015-061b}. Theoretical predictions by \citet{dziem-2016} suggested that these modes may be high-order, $\ell=8$, 9 non-radial modes, which can be calculated from linear pulsation models. Unlike modeling radial modes in the non-linear regime, fitting radial and non-radial modes together using linear models is a technique very similar to the asteroseismology of red giant stars. This method was developed by \citet{netzel2022}, and was further explored, incorporating other observables such as \teff{}, \textit{L} and [Fe/H] information into the fit in \citet{Netzel-2023}.

But, as we have shown, we still lack a way to test the validity of these new RR Lyrae masses through direct methods. We therefore searched for alternatives. RR Lyrae stars make up only a small portion of the He-core-burning regime, and stars exist blue- and redward of them on the HB. Indeed, the evolution of HB stars and their positions on the HB are largely defined by the mass of their H-rich envelopes, as well as the mass loss they experience before reaching the HB. At temperatures below the red edge of the instability strip, we find the red horizontal branch (rHB) stars. These stars are very similar to the RR Lyrae pulsators, except for hosting a slightly more extended envelope. Since their envelopes are also convective, these stars show solar-like oscillations instead of pulsations, and these solar-like oscillations can be analysed in a way very similar to that of red giant and red clump stars \citep{Matteuzzi-2023}. If rHB and RR~Lyrae stars can be observed and modeled within the same population of stars, such as in globular clusters, they can potentially offer a way to compare asteroseismic masses determined by independent methods. 

This opportunity came with the \textit{Kepler} space telescope. \textit{Kepler} observed multiple globular clusters during the K2 mission, including the cluster closest to the Sun, Messier 4 (M4). This cluster was bright enough that seismic data was obtainable not only for the bright red giants, but fainter stars, such as the rHB stars in it, as well \citep{Miglio-2016,Tailo-2022}. \citet{Howell-2022} calculated seismic masses for several stars and compared the averages of various evolutionary phases to estimate the integrated mass loss between stages in the cluster. M4 also contains several RR Lyrae stars, including overtone ones that have been observed by \textit{Kepler}. If non-radial modes can be detected in the overtone stars, their seismic masses could be compared directly to the seismic masses of rHB stars, validating the results.

In Section \ref{sect:data}, we process the Kepler data of five RRc stars observed in M4 and determine their physical parameters. In Section~\ref{sect:rrc-results}, we analyse the photometry and the frequency content of the RRc stars and fit them with seismic models. In Section~\ref{sect:param}, we compare the RRc physical parameters to the (updated) parameters of the red giant stars of the cluster, as well as to theoretical mass relations. Finally, we summarize our findings in Section~\ref{sect:concl}.

\section{Data and methods}
\label{sect:data}
The \textit{Kepler} space telescope observed M4 during Campaign 2 of the K2 mission in late 2014 for 80 days nearly continuously. The cluster was at the edge of module 12 of the detector; hence M4 was only partially covered by the campaign.

\subsection{K2 photometry}
A large portion of the cluster was covered with an extended custom pixel aperture. We identified three RRc stars within this aperture: M4~V6, V42 and V61. To extract their light curves, we followed the same differential-image method based on the FITSH photometry package of \citet{fitsh} that we used on M80 \citep{Molnar-2023}, and on other crowded, faint, and/or moving targets \citep[see, e.g.,][]{Molnar-2015,Molnar-2018,Kalup-2021}. The output for these stars is flux variations relative to the subtracted master frame. We shifted the median variations to the \textit{Gaia} DR3 G-band average brightnesses \citep{GaiaEDR3-2021,Riello-2021}.  

Since M4 has a large angular diameter, it extends beyond the K2 custom aperture. We located two additional RRc stars farther away from the center (V43 and V76), on the outskirts of the cluster: these were recorded with individual pixel apertures. For these two stars, we used the simple aperture photometry (SAP) light curves provided by the mission.  

Raw K2 photometric data contains systematic signals due to the limited pointing capabilities experienced by the telescope during the mission. It was found in earlier works focusing on RR~Lyrae stars \citep{plachy-2019,bodi-2022,Molnar-2023} that the best method to separate the instrumental signals from the pulsations is the K2 Systematics Correction (K2SC) method developed by \citet{K2SC-2016}. We applied the same corrections here to remove fast systematics. Slow trends were then removed with the algorithm developed by \citet{bodi-2022}, fitting a high-order polynomial to the light curve that is optimized for best overlap of the pulsation cycles via phase dispersion minimization. The resulting light curves are shown in Fig.~\ref{fig:lcs}. Photometric data is available in Appendix~\ref{sect:appx}.

One more target, M4~V75, which was identified as a possible RRc star by \citet{Yao-1988}, is located within the K2 custom aperture as well. This star, however, shows no periodic variation, confirming previous non-detections by \citet{Stetson-2014} and \citet{safonova-2016}. 

        \begin{figure}
   \centering
   \includegraphics[width=1.0\columnwidth]{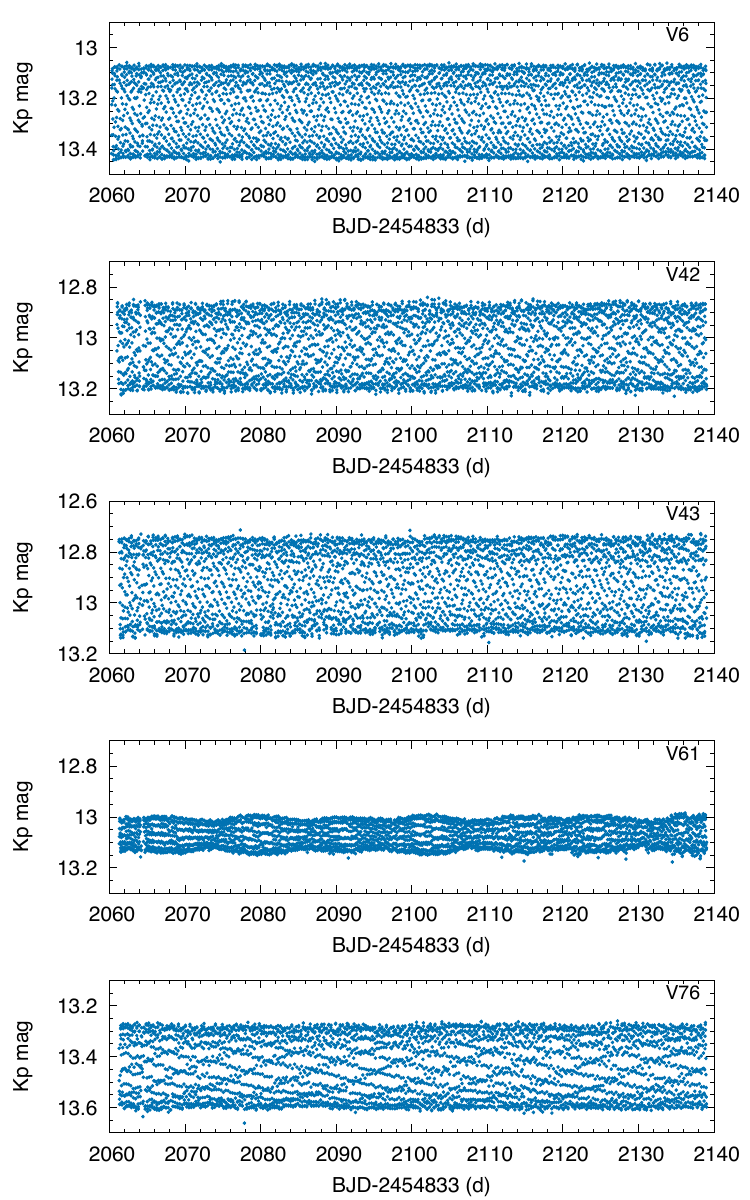}
   \caption{Corrected K2 light curves of the five RRc stars that were targeted by the mission. }
              \label{fig:lcs}%
    \end{figure}

\subsection{Observational constraints}
\label{sect:obs}
A detailed asteroseismic analysis requires further constraints on the physical parameters (such as \textit{L}, \teff{}, $\log g$ or [Fe/H]) of the stars. \citet{Howell-2022} used scaling relations for the red giants in the cluster, which relies on luminosities and effective temperatures. These quantities can be calculated from photometry, but they require accurate corrections for interstellar extinction. Here we used the reddening map of \citet{Alonso-Garcia-2012}, but with the reddening zero point \textit{(E(B--V)}$_0$ = 0.37 mag) determined specifically for M4 by \citet{Hendricks-2012}.

In contrast, our approach for the RRc stars is akin to peak-bagging, where we collect and fit individual oscillation peaks \citep{appourchaux2003}. However, since we are limited to very few modes, it is still important to constrain our modeling space with classical observables such as allowed luminosity, \teff{} and metallicity ranges.   
     \begin{figure}
   \centering
   \includegraphics[width=1.0\columnwidth]{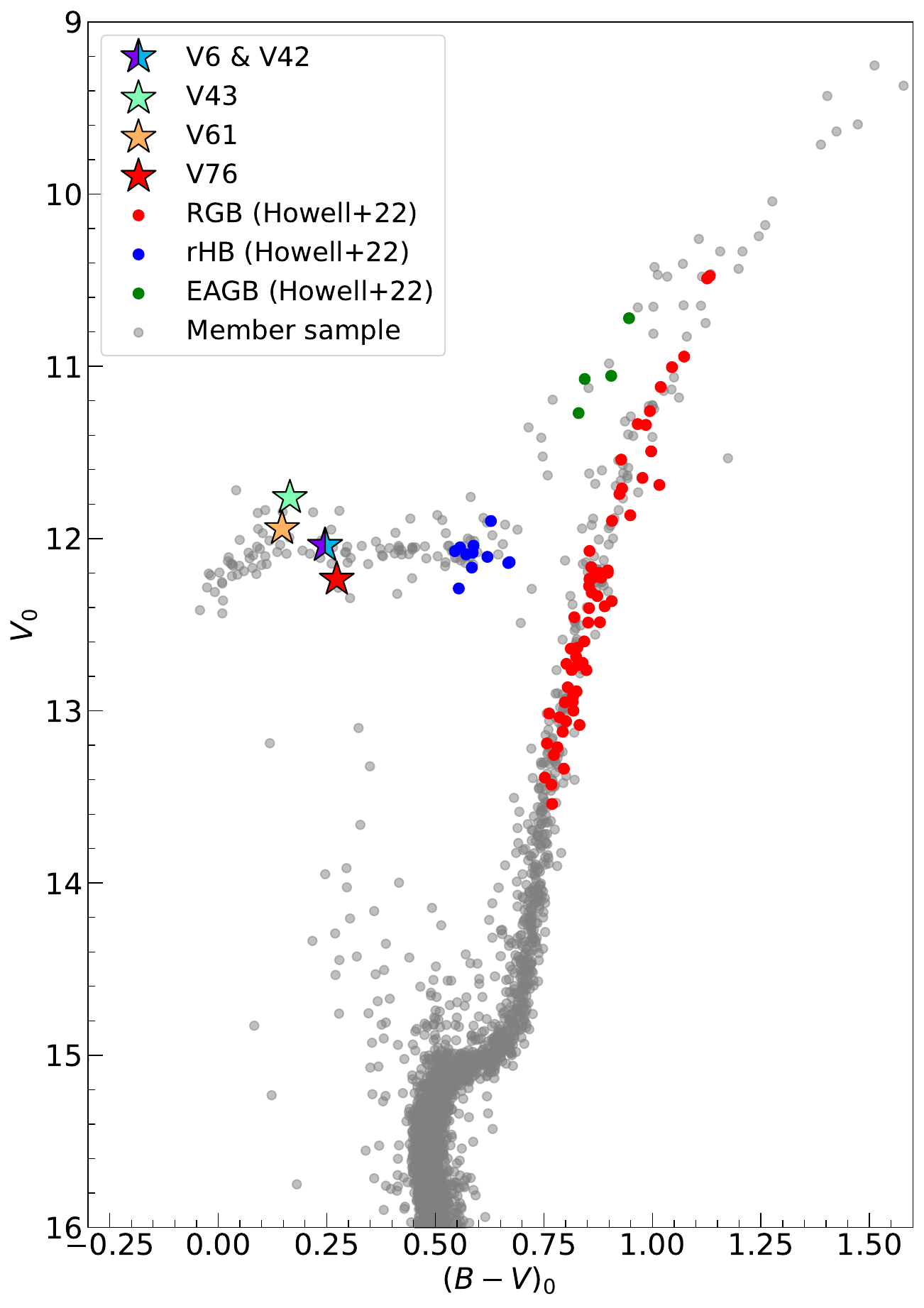}
   \caption{Color-magnitude diagram of M4 in Johnson passbands using photometry from \citet{Stetson-2014,stetson-2019}. Stars targeted by our study are identified by the colored star symbols, and the stars in \citet{Howell-2022} are shown as colored points. V6 \& V42 overlap in the plot, which is indicated by the dual colored marker. The photometry has been corrected for dust using the \citet{Alonso-Garcia-2012} and \citet{pancino-2024} maps. The \textit{Gaia} membership sample (grey) is from \citet{Vasiliev-2021_GalacticGC_memberships}.}
              \label{fig:Johnson-CMD}%
    \end{figure}

We collected the average brightness values of the RRc stars both in \textit{Gaia} DR3 passbands, and in the Johnson \textit{B} and \textit{V} passbands from the observations of \citet{Stetson-2014,stetson-2019}\footnote{Data available at \url{https://www.cadc.hia.nrc.gc.ca/en/community/STETSON/homogeneous/} and at \url{https://www.canfar.net/storage/list/STETSON/homogeneous/}.}. The Johnson data set is displayed in Fig.~\ref{fig:Johnson-CMD}, with the RRc stars and the stars studied by \citet{Howell-2022} highlighted. 
Although former studies by \citet{netzel2022,Netzel-2023} used  \textit{Gaia} data, here we decided to use on the Johnson brightness for consistency with the red giant analysis.

We calculated the insterstellar extinction for the RRc targets similarly to that of the red giant targets, initially relying on the \citet{Alonso-Garcia-2012} and \citet{pancino-2024} differential reddening maps and the same zero point. The latter covers a larger area, but otherwise we found the two maps to be consistent with each other, with differences not exceeding $\pm0.02$ mag. We determined extinctions for four out of five RRc stars this way. We used the zero point of \citet{Hendricks-2012} here, too.  

The fifth star, V43, falls outside each map. For this star, we compared the \textit{E(B--V)} reddening values to the \textit{E(BP--RP)} reddenings from  \textit{Gaia} DR3. We limited the comparison to sources with $G<15$~mag, \textit{(BP--RP)} < 1.4~mag, and E\textit{(BP--RP)} < 1.3~mag, and cut out the core of the cluster. These filters were necessary to remove sources with very high reddenings in the catalog. This local comparison of 67 stars resulted in the conversion of $E(B-V) = 0.169\,E(BP-RP) + 0.290$. We note, however, that this formula has not been tested more extensively, and such conversions elsewhere will require further calibrations. Furthermore, the RR Lyrae variables themselves have \textit{Gaia} reddenings considerably larger than their neighboring stars, so we instead calculated the median of nine neighboring stars for V43.   

We then used the distance modulus provided by \citet{Baumgardt-2021}, $\mu = 11.337 \pm 0.018$~mag, to convert the extinction-corrected values into absolute magnitudes.

       \begin{figure*}
   \centering
   \includegraphics[width=1.0\textwidth]{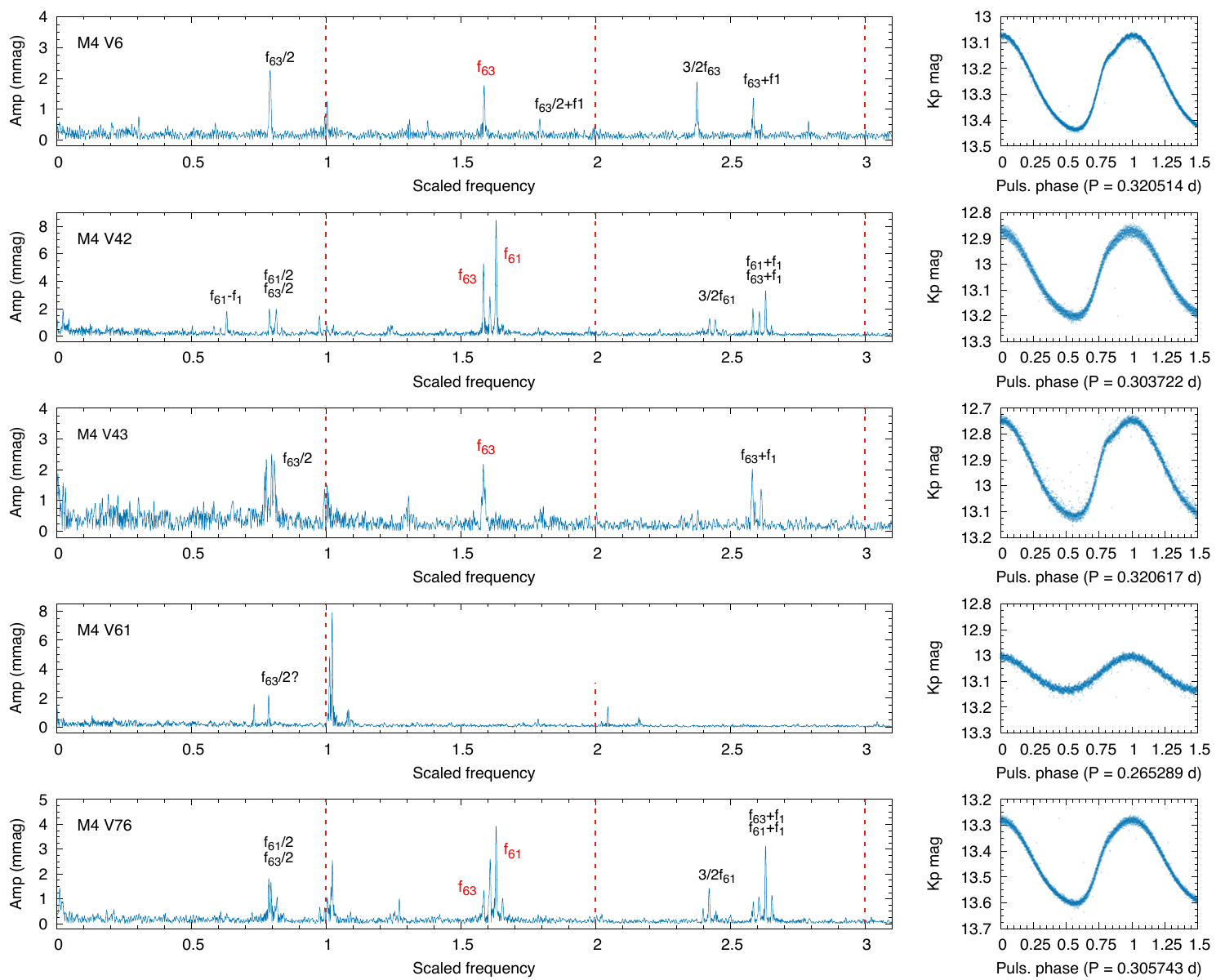}
   \caption{Left: Fourier spectra of the five RRc stars. Here we removed the main pulsation frequency and its harmonics belonging to the first overtone to reveal the low-amplitude extra modes. Modes are labeled with red, and combination frequencies with black. Right: light curves folded with the first overtone period.}
              \label{fig:four-spectra}%
    \end{figure*}

The effective temperature, $T_{\rm eff}$, was calculated using the color-$T_{\rm eff}$ relation by \citet{gonzalezhernandez.bonifacio2009} with (V-K)$_0$ colors. For four stars, we used the $(V-K)_0$ color indices using $K$-band photometry from 2MASS \citep{2mass}. V43 lacks $K$-band photometry: for that star we set the range of $\log T_{\rm eff}$ to $3.80 - 3.90$ dex. The dereddened color $(V-K)_0$ was calculated using the color excess $E(V-K)$ obtained from $E(B-V)$ with the transformation relation of \cite{fitzpatrick.massa2007}. The calculated ranges of $T_{\rm eff}$ are collected in Table~\ref{tab:input_values}.

We set the metallicity to [Fe/H] $\approx -1.1 \pm 0.07$ \citep{Harris-1996, Marino-2008, MacLean-2018}. The luminosity was calculated using the formula:

\begin{equation}
    L/L_{\rm \odot} = 10^{(-0.4\,(M_{\rm bol}-(M_{\rm bol}^\odot))},
\end{equation}

where $M_{\rm bol}$ is a bolometric brightness of a star, and $M_{\rm bol}^\odot$ is a bolometric brightness of the Sun, set to 4.74\,mag. We converted M$_V$ to $M_{\rm bol}$ using the bolometric correction calculated with the relation by \cite{alonso1999}, which uses $T_{\rm eff}$ and metallicity [Fe/H]. The resulting luminostiy ranges are collected in Table~\ref{tab:input_values}.

Metal, $Z$, and helium, $Y$, contents were calculated based on [Fe/H]. First, we transformed [Fe/H] to [M/H] using formula by \cite{Salaris-1993}:

\begin{equation}
    {\rm [M/H]} \approx {\rm [Fe/H]} + \log \left(0.638 \cdot 10^{[\alpha/{\rm Fe}]} + 0.362 \right),
\end{equation}
where [$\alpha$/Fe] is enhancement by $\alpha$ elements. Metal-poor populations have elevated $\alpha$ element abundances: for M4, \citet{Marino-2008} found [$\alpha$/Fe]\,=\,$+0.39\pm0.05$ \citep{Johnson-2014,Bensby-2017}. 

Then, we transformed [M/H] to $Z$, using the relation:

\begin{equation}
    {\rm [M/H]} = \log\frac{Z}{\rm Z_\odot} - \log\frac{X}{\rm X_\odot},
\end{equation}
where $\rm Z_\odot$ and  $\rm X_\odot$ are solar values, which we set to $\rm Z_\odot = 0.0134$ and  $\rm X_\odot = 0.7381$ \citep{Asplund-2009}.

Assuming that hydrogen content, $X$, can vary from 0.70 to 0.76, and $\alpha$-element enhancement can vary from 0 to 0.4, we calculated the modeling range for $Z$ to be 0.00086 -- 0.0032.

\section{Photometric results on the RRc stars}
\label{sect:rrc-results}
We used the \texttt{Period04} software to determine the frequency composition of each star \citep{period04}. The detection limit was set to S/N > 4.0, relative to the nearby average noise level in the frequency spectrum. First, we subtracted the pulsation frequency and its harmonics, then searched for any extra signals above or below the pulsation frequency. Folded light curves and frequency spectra are displayed in Fig.~\ref{fig:four-spectra}.

\subsection{Extra modes}

According to \citet{dziem-2016}, the signals around period ratios $P/P_{\rm O1} \approx 0.63 $ and 0.61, relative to the first overtone, correspond to (the first harmonic of) $l=8$ and $l=9$ modes, respectively. Here we label these as $f_{63}$ and $f_{61}$ frequencies, although this family of modes is also called $f_{61}$ or $f_{X}$ modes collectively. The actual mode frequencies are the subharmonics of these signals at $f_{61}/2$ and $f_{63}/2$, but those are usually harder to detect due to geometric cancellation effects.

We clearly identified the modes in four stars: V6, V42, V43 and V76. We also see subtle differences between them. In V6, the subharmonic at the true pulsation frequency is the strongest, which is unusual among RRc stars as we expect the harmonic to be stronger. In V42 and V76, we detect multiple peaks and the corresponding harmonics and subharmonics. In V43, the peak appears to be incoherent and can be fitted with three close-by frequencies. Here we observe a very wide structure at the subharmonic that extends below the expected range of the $f_{61}$ modes. However, these frequencies are not low enough to be either the fundamental mode or an $f_{68}$ mode.   

While $f_{61}$ modes are prevalent in the RRc population of M4, we did not detect the low-frequency $f_{68}$ mode clearly in either of the targets. The latter is a separate group of extra modes, discovered recently in RRc stars, which appear at lower frequencies $(f/f_{\rm O1} \approx 0.686)$, and whose origin has not been identified yet \citep{netzel2015,Benko-2023}. The lack of detection is in agreement with the mode abundance results of\linebreak \citet{netzel-2024}. They found that while the $f_{61}$ modes are most frequent around [Fe/H]~$\approx -1.0$, the frequency of the $f_{68}$ mode drops drastically above -1.3, and the metallicity of M4 is higher than that limit.

In contrast to the results above, V61 turned out to be the rather different from the four other stars. Here we only detect significant extra frequency peaks at the subharmonic range, but not at the expected $f_{61}$ range. The two peaks appear at period ratios 0.7335 and 0.7877. The former value may potentially correspond to the fundamental mode, with a period of 0.36168~d: however, that would put the star to [Fe/H] values higher than that of M4 in the Petersen diagram \citep{szabo-2004,Chen-2023}. We investigate the modulation properties of the star in Appendix~\ref{sec:v61}. 

\begin{figure}
   \centering
   \includegraphics[width=1.0\columnwidth]{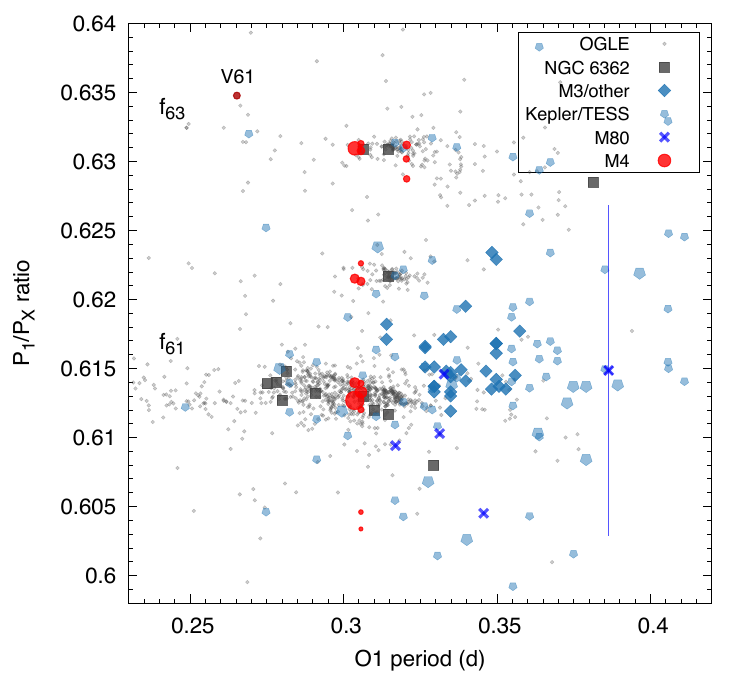}
   \caption{Petersen diagram of the $f_{61}$-type modes. Here we plot signals detected in M4 (in red) over existing literature data. Frequency peaks fall onto the main ridges discovered in the OGLE data. Positions of the $f_{61}$ and $f_{63}$ ridges are labelled respectively.}
              \label{fig:petersen}%
    \end{figure}

\subsection{Petersen diagram}

We present the identified $f_{61}$ signals on a Petersen diagram in Fig.~\ref{fig:petersen}, showing the period ratios against the longer period (here fixed to the first overtone).  We plot the M4 results in red against the points presented by \citet{Molnar-2023}. We find that the points line up well with other samples that have moderate metallicities, such as the bulge and NGC~6362. In this regime most frequencies clearly fall onto the main ridges at period ratios $P_1/P_X \approx 0.632$ and 0.613, corresponding to the $\ell=8$ and 9 modes. We also detect signals in the middle ridge, which \citet{dziem-2016} interprets as combination frequencies. Further signals appear below, at $P_1/P_X \approx 0.605$ which might also be combination peaks. We did not detect any $\ell=10$ peaks which should be at $P_1/P_X \approx 0.6$ or below, according to \citet{dziem-2016}.

\subsection{Comparison with other photometric results}

The K2 observations were analysed before by \citet{wallace-2019} who searched the cluster for new variables. They detected higher scatter in V61 than in other RR Lyrae stars, but did not recognize it as modulated. They also detected two stars they classified as millimagnitude RR Lyrae stars \citep{wallace-mmRRL-2019}. Both of those two stars show only a single periodicity.

Ground-based multicolor photometry was collected for the RR Lyrae members (among others) by \citet{Stetson-2014}. They did not recognize V61 as a modulated star either, although upon reanalysis of their photometry, a side peak is visible. The only other star common with our targets that has a useful amount of time series photometry is V6. There a frequency signal is marginally detectable at the position of the subharmonic ($f_{63}/2$). This shows that seismic modeling of RRc stars requires extensive, high-precision photometry to detect the low-amplitude extra modes.

We also analyzed the observations collected by the Zwicky Transient Facility \citep[ZTF,][]{ZTF-2019} and the Asteroid Terrestrial-impact Last Alert System \citep[ATLAS,][]{ATLAS-2018}. The ZTF data was too sparse for a detailed frequency analysis. The ATLAS survey collected significantly more data points, but the photometric precision of the observations was too low to detect any extra modes.

\begin{table*}
    \centering
    \caption{Input parameter ranges, pulsation periods and period ratios for each RRc star that we used for modeling. PR1 and PR2 are the mode period ratios relative to the first overtone. Mass ranges are left deliberately wide to accommodate realistic values over the grid.}
    \label{tab:input_values}
    \begin{tabular}{c|ccccc}
    \hline
    \hline
         ID & V6 & V42 & V43 & V61 & V76 \\
         \hline
         M $(M_\odot)$ & $0.5-0.9$ & $0.5-0.9$ & $0.5-0.9$ & $0.5-0.9$ & $0.5-0.9$ \\
         log$L\,( L_\odot)$ & 1.59 -- 1.68 & 1.59 -- 1.68 & 1.67 -- 1.81 & 1.63 -- 1.73 & 1.51 -- 1.60 \\
         log$T_{\rm eff}$ & 3.81 -- 3.86 & 3.81 -- 3.86 & 3.80 -- 3.90 & 3.86 -- 3.93 & 3.77 -- 3.83 \\
         $X$ & 0.7 -- 0.76 & 0.7 -- 0.76 & 0.7 -- 0.76 & 0.7 -- 0.76 & 0.7 -- 0.76 \\
         $Z$  & 0.00086 -- 0.0032 & 0.00086 -- 0.0032  & 0.00086 -- 0.0032 & 0.00086 -- 0.0032 & 0.00086 -- 0.0032 \\ 
         $P_{\rm 1O}$ (d) & 0.320506 & 0.303719 & 0.320622 & 0.265290 & 0.305737 \\
         PR1  & 0.63005 & 0.63095 & 0.63106 & 0.63476 & 0.63076 \\
         PR2  & -- & 0.61268 & -- & -- & 0.61327 \\
        \hline
    \end{tabular}
\end{table*}

\subsection{RRc pulsation models}

For each star, we calculated pulsation models to match the observed first-overtone period and period ratio(s). To calculate the theoretical models, we used the envelope pulsation code of \cite{Dziembowski-1977}. The input physical parameters required by the code are mass, luminosity, effective temperature, hydrogen and metal abundances. Additionally, this code requires an approximate value of dimensionless frequency for the non-radial mode. We followed the approach of \cite{netzel2022} for individual calculations. Namely, we used the estimates that relate non-radial mode frequencies from the linear fits to sequences in the Petersen diagram \citep[see Equations 4 and 5 in][]{netzel2022}, which were then converted to dimensionless frequencies, $\sigma$, using the formula from Eq.~6 in \cite{netzel2022}. Furthermore, we set the range of possible starting values of non-radial mode frequencies to $\sigma \pm 0.1$. Consequently, for each model we were able to get the non-radial mode with the highest driving rate, i.e., it is the most strongly trapped in the envelope. 

In order to find the best matches between periods and period ratios in theoretical models and in observed values for each star, we employed genetic algorithms\footnote{We used a Python \texttt{geneticlagorithm} library \url{https://github.com/rmsolgi/geneticalgorithm}}, which used the  pulsation code with different input physical parameters. The ranges of physical parameters were set based on observational constraints (see Table~\ref{tab:input_values}). In the case of mass, we set a wide range of $0.5 - 0.9$ M$_\odot$ to accommodate any realistic mass result. For each star, we executed 50 separate runs. For each run we set the maximum number of iterations to 150, population size of 100, mutation probability of 0.1, elitist ratio to 0.01, crossover probability to 0.5, and parents portion to 0.3. We note that we performed calculations with different parameters for genetic algorithms beforehand, including the numbers of populations and iterations, to ensure that the results are robust. The values and errors of physical parameters were derived as the means and standard deviations of the results from each of the fifty individual runs.

Calculated ranges of physical parameters used to constrain the models for each star are summarized in Table~\ref{tab:input_values}. Correlations between individual parameters based on the distributions of the results from each run are visualized with corner plots and discussed in Appendix~\ref{sect:corner}.

  \begin{figure}
   \centering
   \includegraphics[width=1.0\columnwidth]{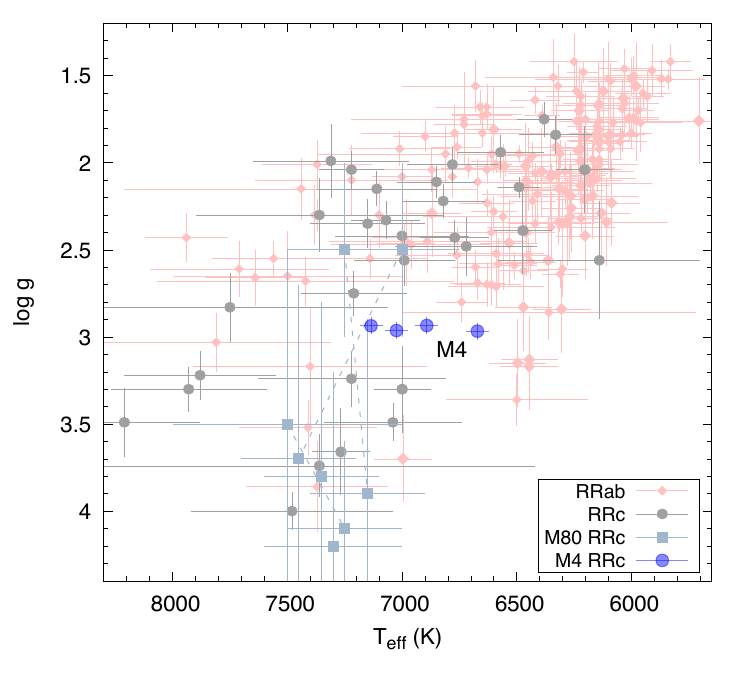}
   \caption{Positions of the seismically determined $T_{\rm eff}$ and $\log\, g$ values for the five RRc stars in M4, against the spectroscopically observed RRc stars in
the $T_{\rm eff}$--$\log g$ plane, as collected by \citet{Molnar-2023}.}
              \label{fig:logg}%
    \end{figure}

\begin{table*}
    \centering
    \caption{Modeled physical parameters of the five RRc stars based on seismic modeling.}
    \label{tab:output_values}
    \begin{tabular}{c|ccccc}
\hline
\hline
ID & V06 &  V42 & V43 & V61 & V76  \\
\hline
M $(M_\odot)$ & $ 0.636 \pm 0.030$ & $0.642 \pm 0.026$ & $0.664 \pm 0.048$ & ($0.705 \pm 0.047$) & $0.667 \pm 0.019$ \\
$\log L\,(L_\odot)$ & $ 1.636 \pm 0.023$ & $1.621 \pm 0.020$ & $1.690 \pm 0.011$ & ($1.689 \pm 0.024$) & $1.545 \pm 0.019$ \\
\teff & $ 6982 \pm 35$ & $7024 \pm 31$ & $7134 \pm 29$ & ($7471 \pm 133$) & $6672 \pm 55$ \\
$\log{}g$ & $ 2.935 \pm 0.007$ & $ 2.963 \pm 0.006 $ & $ 2.936 \pm 0.008 $ & $ (3.043 \pm 0.007) $ & $ 2.967 \pm 0.003 $ \\
X & $ 0.731 \pm 0.020$ & $0.735 \pm 0.018$ & $0.731 \pm 0.020$ & ($0.725 \pm 0.018$) & $0.742 \pm 0.017$ \\
Z & $ 0.0019 \pm 0.0007$ & $0.0015 \pm 0.0004$ & $0.0016 \pm 0.0006$ & ($0.0022 \pm 0.0006$) & $0.0011 \pm 0.0002$ \\
\hline
    \end{tabular}
\end{table*}

\section{New and revised physical parameters}
\label{sect:param}
With the RRc physical parameters obtained, we proceeded to compare them to parameters of other RR Lyrae stars, as well as to the average properties of M4. For further comparison, we recalculated the masses for the oscillating rHB stars originally studied in \citet{Howell-2022}, too.

\subsection{RR Lyrae masses and other physical parameters}
As listed in Table~\ref{tab:output_values}, the masses of four out of the five RRc stars are within 0.03~$M_\odot$, between 0.636 and 0.667~$M_\odot$ with an average uncertainty of $\pm 0.031~M_\odot$. These masses align well with the mass-period distribution of RRc stars published by \citet{Netzel-2023}.

The only outlier is the modulated star, V61, for which we found no solutions with unstable non-radial modes. This is caused by the higher \teff{} value of the star which puts it near the blue edge of the RR~Lyrae instability strip, an outside of the instability region of the 0.61 modes \citep{netzel2022}. This suggests that the frequency component we detected and identified tentatively as $f_{63}/2$ might have a different origin. If we allow for solutions with stable modes at the detected mode frequency, we find a mass of $0.71\pm 0.05\, {\rm M}_\odot$ which is significantly higher than the rest. We include this value for the sake of completeness, but consider it as a marginal detection and do not include it in our statistics. 

We compared the calculated $T_{\rm eff}$ and $\log\,g$ values to the distribution of RR Lyrae stars presented by \citet{Molnar-2023}. As Fig.~\ref{fig:logg} shows, the stars belong to the cooler RRc stars, lying between 6600--7200~K, with $\log\,g$ values being very close to 3.0 for all five stars. For chemical composition, we find average bulk abundances of $X = 0.734 \pm 0.014$ and $Z = 0.0016 \pm 0.0005$. These values indicate a He abundance of $Y = 0.264 \pm 0.014$ for M4, pointing towards low He enrichment ($\Delta Y<0.02$) relative to the primordial He abundance, which is in tension with the Y values inferred by \citet{Valcarce-2014} spectroscopically. On the other hand, our result is in agreement with the model-based inferences of \citet{Villanova-2012}, supporting their conclusion that the spectroscopic data should be reevaluated, and followed upon by further observations. We note, however, that the distribution of X values among the individual model runs are skewed towards higher values for two stars (V42, V76), but are bimodal for the other two (V06 and V43) as shown in Appendix~\ref{sect:corner}. For V61 the fits skew towards lower values but since it is a marginal detection we do not consider it in the averages. 

The comparison of the mass ratio of heavy elements, \textit{Z}, to the observed [Fe/H] index is not straightforward, as it is influenced by the abundance differences of individual elements \citep[see, e.g.,][]{Hinkel-2022}. Nevertheless, we can give an estimate for the correctness of our seismic \textit{Z} values. Following the relations described in Section~\ref{sect:obs}, in reverse order, and using an $\alpha$-enhancement of [$\alpha$/Fe] = $0.39 \pm 0.05$ \citep{Marino-2008}, we get an average seismic [Fe/H] of $-1.13\pm 0.05$ for the five RRc stars, which agrees with the spectroscopic [Fe/H] value of the cluster. 

\subsection{Seismic masses for red giants}
Mixed mode oscillations in red giants are characterized by two global seismic parameters; the large frequency separation, $\Delta\nu$, and the frequency of the maximum acoustic power, $\nu_{\text{max}}$. 
These quantities are correlated to stellar properties, which are used to derive seismic mass scaling relations \citep{Ulrich-1986,Brown-1991,kjeldsen-1995}. For faint globular cluster stars, the short observing baseline of the K2 mission results in lower frequency resolution and signal-to-noise, and as such it can be difficult to estimate $\Delta\nu$. However, accurate asteroseismic masses can be measured independently of a $\Delta\nu$ estimate \citep[e.g.][]{Howell-2024,Howell-2025, Reyes-2025_M67}, using the following scaling relation:
\begin{align}
    \label{eq:mass_relation3}
    &\left(\frac{M}{M_{\odot}}\right)\simeq\left(\frac{\nu_{\text{max}}}{\nu_{\text{max},\odot}}\right)\left(\frac{L}{L_{\odot}}\right)\left(\frac{T_{\text{eff}}}{T_{\text{eff},\odot}}\right)^{-7/2}.
\end{align}

We use the oscillating red giant sample and the corresponding K2 light curves from \citet{Howell-2022} to measure seismic masses using Eq.~\ref{eq:mass_relation3}. They detected solar-like oscillations in 75 red giants across three phases of evolution (RGB, rHB and early AGB), and used the \texttt{pySYD} pipeline \citep{Chontos-2021} to measure the asteroseismic parameters. We adopt the \citet{Howell-2022} estimates for $T_{\text{eff}}$ and luminosity, although not their $\nu_{\text{max}}$ measurements. Here we report updated measurements of $\nu_{\text{max}}$ and the corresponding uncertainty for each star using a new asteroseismic pipeline, \texttt{pyMON} \citep{Howell-2025}.

The \texttt{pyMON} pipeline is an adaptation of \texttt{pySYD} (see \citealt{Chontos-2021} for more details), where the main difference is that $\Delta\nu$ is not measured. In both pipelines, the power spectrum is smoothed using an estimate for $\Delta\nu$, however \texttt{pyMON} uses a calibrated $\Delta\nu$-$\nu_{\text{max}}$ scaling relation \citep{Stello-2009_dnu_numax_relation} to estimate this quantity. This is useful for low signal-to-noise data -- such as the M4 red giant sample in \citet{Howell-2022}-- where measurements are $\Delta\nu$ are highly uncertain. A comparison between the \texttt{pySYD} results in \citet{Howell-2022} and our \texttt{pyMON} results show that the measurement of $\nu_{\text{max}}$ remains consistent, however there is a reduction by a factor of $\sim2$ in the $\nu_{\text{max}}$ uncertainty (refer to \citealt{Howell-2025} for further discussion on this uncertainty reduction). Consequently, this results in a decrease in the seismic mass uncertainties by a factor of $\sim1.5$. 

Scaling relations may also include correction factors ($f_{\nu{\rm max}}$, $f_{\Delta\nu}$) to account for structural differences between the Sun and the target stars. In this case we are only interested in the value of $f_{\nu{\rm max}}$, for which various prescriptions have been given in the literature \citep[see, e.g.,][]{Sharma-2016,Hekker-2020,Pinsonneault-2025}, but these have been limited to higher metallicities. More recently, \citet{Lindsay-2025} found that at the metallicity of M4, $f_{\nu{\rm max}}\approx1.05$, which could indicate an overestimation of about 5\% of RGB masses. However, that only applies to RGB stars, and correction factors have not been estimated for low-metallicity core-He-burning or AGB stars yet. 

We note, however, that correction factors are predominantly calibrated through (or involving) stellar radii, and \citet{Ash-2025} showed that radius-based $f_{\nu{\rm max}}$ values could potentially result in overcorrecting the masses. They also found that while not using any $f_{\nu{\rm max}}$ corrections can result in small systematic errors in the seismic masses, the differences stay largely constant along the RGB (and larger, $\nu_{\rm max}$-dependent deviations only occur when $\Delta\nu$ is also used). Thus, if $f_{\nu{\rm max}}$ does not depend on $\nu_{\rm max}$, the mass differences between different stars will not be affected. 

Furthermore, clusters offer the possibility to test the masses against stellar models and isochrones. \citet{Howell-2022} found that the RGB masses agree with the masses of the fitted isochrone without the need for correction factors. We also show the agreement between our RGB masses and stellar models in Sect.~\ref{sect:HRD}. We also note that decreasing our AGB masses by 5\% would be in tension with the white dwarf masses in the cluster (see Sect.~\ref{sect:avgmasses}). Based on these arguments, and also to stay consistent with analyses of other globular clusters which used $f_{\nu{\rm max}}=1$ \citep{Howell-2022,Howell-2024,Howell-2025}, we decided to keep $f_{\nu{\rm max}}$ at unity. 

  \begin{figure}
   \centering
   \includegraphics[width=1.0\columnwidth]{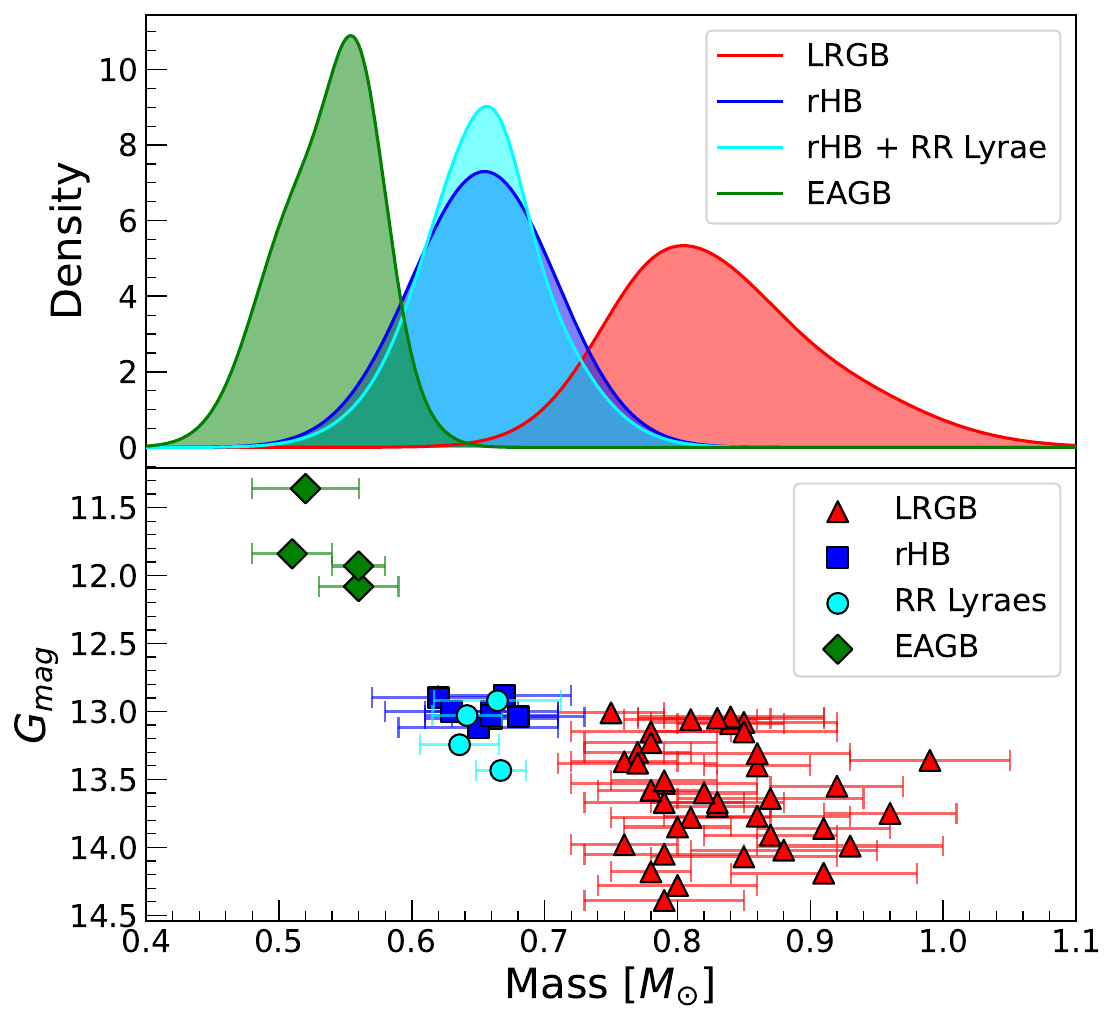}
   \caption{Top: The mass distributions (calculated as KDEs) for the M4 sample of red giants and RR Lyraes, separated into evolutionary phases. For the HB sample, we have included a KDE consisting of just the rHB sample (blue), and a KDE which combines both the rHB and RR Lyrae samples (cyan). We note that the area of each KDE is normalised to one, and as such the heights of the distributions do not correspond to sample sizes. Mass outliers have been removed. Bottom: Individual masses for the red giant and RR Lyrae samples plotted against \textit{Gaia} magnitude.}
              \label{FigGam}%
    \end{figure}

\subsection{Average masses of the subpopulations}
\label{sect:avgmasses}
Most Galactic globular clusters contain multiple populations of stars, usually signaling two consecutive waves of star formation \citep{Milone-2022}. M4 is no exception to this: the possible signs of both populations in the RGB data and their mass difference was already explored by \citet{Howell-2022}. However, since the RGB population showed only weak signs of bimodality at the mass, disentangling the RGB population will require further chemical abundance measurements. Since here we are more concerned about the later stages of stellar evolution, we treat the RGB sample as a single population. Using spectroscopy, \citet{MacLean-2018} reported that all AGB stars in M4 belong to the first generation of stars, with second-population stars possibly skipping the AGB altogether. This suggests that the EAGB stars in \citet{Howell-2022} only belong to the first generation. Most importantly, \citet{Marino-2011} found that both the rHB and RR Lyrae stars of the cluster are first-population stars, and the distinct blue arm of the HB contains the second population. This makes it possible to compare their masses to each other and to the EAGB stars, too.  

\begin{figure}
   \centering
   \includegraphics[width=1.0\columnwidth]{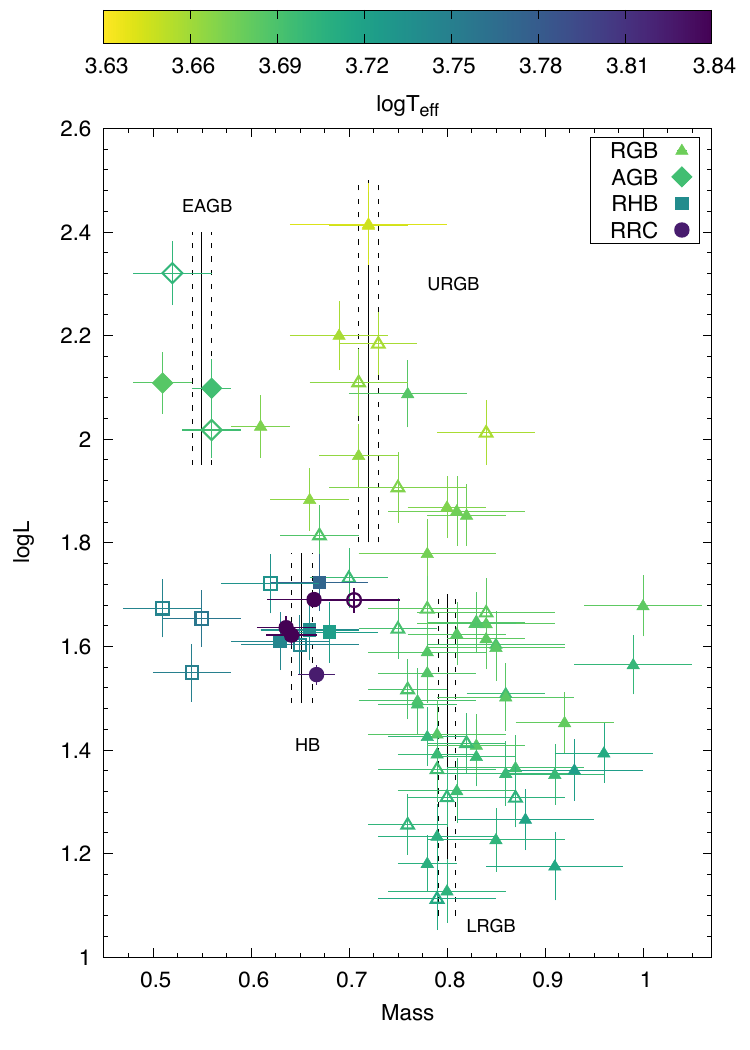}
   \caption{The distribution of masses with increasing luminosity among the evolutionary groups. Lines indicate the average masses and uncertainties.
   }
              \label{fig:massvlum}%
    \end{figure}

In Figure~\ref{FigGam}, we show the distribution of our masses as kernel density estimation (KDE) functions (top panel) and as a scatter plot (bottom panel) for each evolutionary phase. We removed the stars identified as mass outliers in \citet{Howell-2022} and our study (V61). Additionally, we limited the RGB sample to below the luminosity bump (hereby known as lower RGB or LRGB) to ensure there was no bias to lower masses due to any mass loss that might have occurred. To estimate a new average mass for the rHB sample, we determine the peak value of the corresponding KDE. This represents the mode mass for this sample, and was measured to be $0.655 \pm 0.007~\mathrm{M}_{\odot}$, where the uncertainty is calculated as the standard error on the mean. Our average rHB mass is consistent with the value found previously in \citet{Howell-2022}. This was similarly found for the RGB and early AGB samples, where we calculated average masses of $0.80\pm0.009~\mathrm{M}_{\odot}$ and $0.55\pm0.01~\mathrm{M}_{\odot}$ respectively\footnote{We highlight that \citet{Howell-2022} used the mean mass as their average mass scale. In this paper, we adopt the same method as \citet{Howell-2024} of measuring the mode mass of the KDE distributions to determine the average mass. Both average mass measures are consistent within uncertainties.}. Individual and average masses relative to the calculated luminosities are plotted in Fig.~\ref{fig:massvlum}, and listed in Table~\ref{tab:masses}. 

\begin{table}
    \centering
    \caption{Average masses of each evolutionary group.}
    \label{tab:masses}
    \begin{tabular}{l c c}
\hline
\hline
Evol.~stage & Average mass ($M_\odot$) &  Uncertainty ($M_\odot$) \\
\hline
Lower RGB & 0.800 & 0.009  \\
Upper RGB & 0.717 & 0.012 \\
HB (all)  & 0.652 & 0.022 \\
red HB    & 0.657 & 0.034 \\
RRc stars & 0.648 & 0.028  \\
early AGB & 0.550 & 0.010  \\
\hline
    \end{tabular}
\end{table}

The integrated mass loss between evolutionary phases are also consistent with \citet{Howell-2022}. The mass loss from the lower RGB to the HB decreases slightly to $\Delta M_{\rm RGB-HB} = 0.145 \pm 0.011$, whereas we find a mass loss from $\Delta M_{\rm HB-AGB} = 0.105 \pm 0.013$ from the HB to the early AGB.

A further mass constraint for the cluster was determined by \citet{Kalirai-2009}, who inferred masses of 0.50--0.55 M$_\odot$ for six white dwarfs. As discussed by \citet{Howell-2022}, this is very close to the AGB seismic masses, indicating very low mass loss on the AGB, which can be explained by the stars skipping most or all of the thermally pulsing stage of the AGB. We note that discrepancies between AGB evolutionary models and elemental abundances in the AGB stars have also been reported \citet{MacLean-2018}. A KDE-based fitting of the WD data, consistent with our analysis of the other populations, resulted in an average mass of $\overline{M}_{\rm WD} = 0.52\pm0.07$\,M$_\odot$, which is lower than the average published by \citet{Kalirai-2009}. While this is still quite close to the EAGB mass, the difference is 0.03\,M$_\odot$, suggesting a small amount of AGB mass loss. We note, however, that the AGB mass distribution is based on only four stars, which makes the AGB average mass rather uncertain.

\subsection{The seismic HRD of M4}
\label{sect:HRD}

\begin{figure}
   \centering
   \includegraphics[width=1.0\columnwidth]{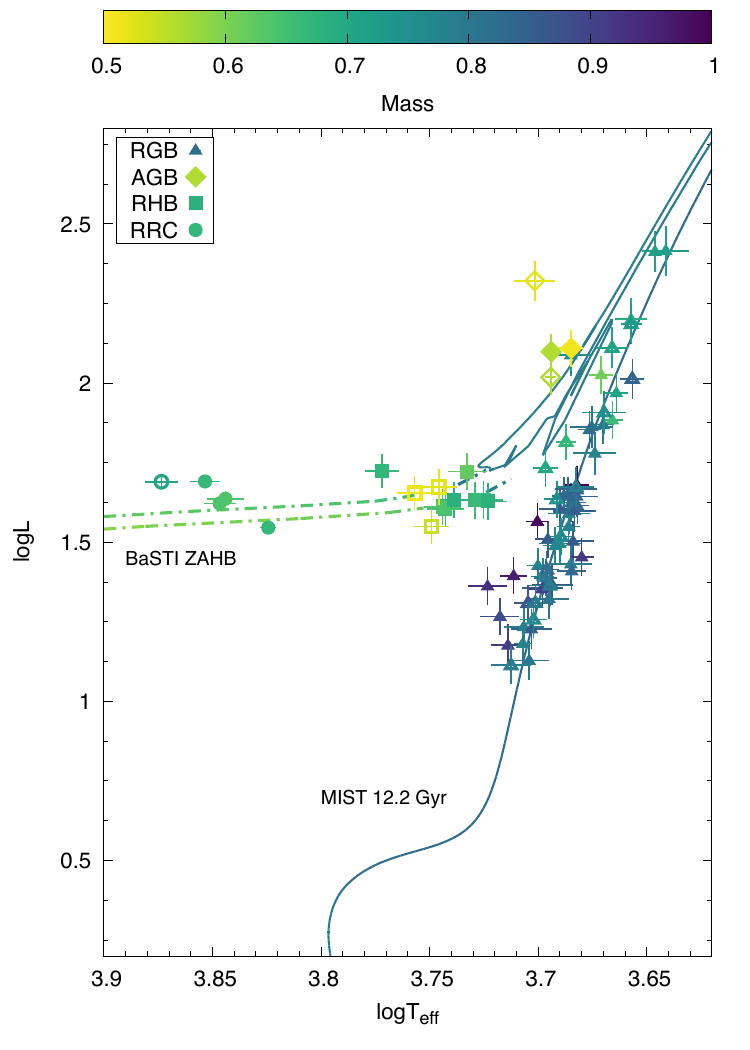}
   \caption{The seismic HRD of M4. Here we show stars in the $\log\,L - T_{\rm eff}$ plane for which masses are available. Colors indicate stellar mass. The gap in the HB between RRc and rHB stars is where RRab stars reside, for which no mass estimate is available. The solid line is a MIST isochrone with an age of 12.2 Gyr and an [Fe/H] index of --0.9. Two BaSTI ZAHB tracks ([$\alpha$/Fe]\,=\,0.4; [Fe/H]\,=\,--0.9 (lower) and [Fe/H]\,=\,--1.2 (upper), respectively) are also shown.} 
              \label{fig:realhrd}%
    \end{figure}

We plot the distributions of the masses along the HRD in Fig.~\ref{fig:realhrd}. We also include a MIST\footnote{\url{https://waps.cfa.harvard.edu/MIST}} (MESA Isochrones and Stellar Tracks; \citealt{MIST-Choi-2016}) isochrone for comparison. This isochrone has an age of 12.2~Gyr, an average of various literature sources for M4 \citep{age-2022}. Since MIST does not offer $\alpha$-enhanced isochrones, we use a value of [Fe/H] = --0.9 to emulate the higher [$\alpha$/Fe] abundances, based on the scaling described by \citet{Joyce-2023}. The mass loss prescriptions used by MIST for low-mass stars are $\eta_R = 0.1$ and $\eta_B=0.2$ for the RGB and AGB phases, following the \citep{Reimers-1975} and \citet{Bloecker-1995} schemes, respectively. Colors represent masses: it is clear that the conservative mass loss prescription used in MIST's pre-computed isochrone database does not capture the true amount of mass loss in the cluster. This is most evident in the later evolutionary stages.

The isochrone has masses around 0.82--0.83~$M_\odot$ in the RGB, which agrees with our inferred LRGB masses to within 3\%, even without further optimization of the isochrone fit. We note that this again indicates that no correction is necessary for the scaling relation, since literature $f_{\nu{\rm max}}$ values would decrease the seismic masses. However, the isochrone mass only drops to 0.79~$M_\odot$ by the time of the HB and early AGB, which is significantly higher than the observed values. A consequence of the overly conservative mass loss prescription is that the isochrone barely reaches the red edge of the rHB region and therefore cannot reproduce the structure of the HB at all. We note that mass loss parameters cannot be adjusted in MIST through the web interface, either. 

Our results indicate that for old, low-mass stellar populations, such as globular clusters, pre-packaged MIST isochrones (and any other isochrone set with similarly conservative mass loss prescriptions) can only reliably reproduce the masses of stars up to the lower RGB, before reaching the red giant branch bump. This highlights the importance of cluster seismology and realistic mass loss estimates, because these studies can provide the first observational constraints on the required level of mass loss through the later stages of low-mass stellar evolution. 

In order to capture the structure of the HB, we also included two zero-age horizontal branch (ZAHB) tracks in Fig.~\ref{fig:realhrd} from the Bag of Stellar Tracks and Isochrones (BaSTI) model library \citep{Basti-2021}. Both tracks use alpha-element enhanced models ([$\alpha$/Fe]\,=\,0.4). The two pre-computed tracks closest to the [Fe/H] index of M4 have indices of --0.9 and --1.2, respectively. We also tested alpha-enhanced BaSTI isochrones at the same age as the MIST one in the plot, but they only differed significantly from MIST at the main sequence turnoff and were nearly indistinguishable along the giant branch.

    \begin{figure*}
   \centering
   \includegraphics[width=0.652\textwidth]{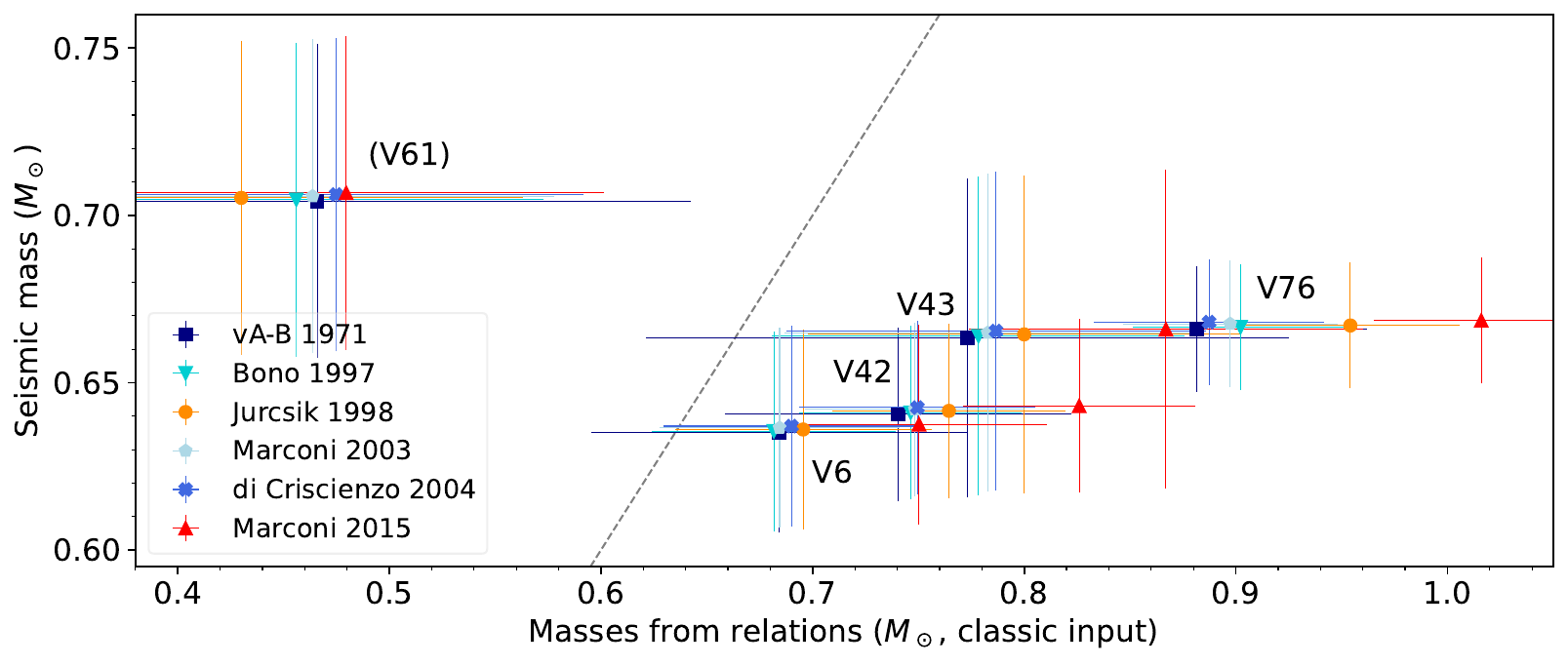}
   \includegraphics[width=0.338\textwidth]{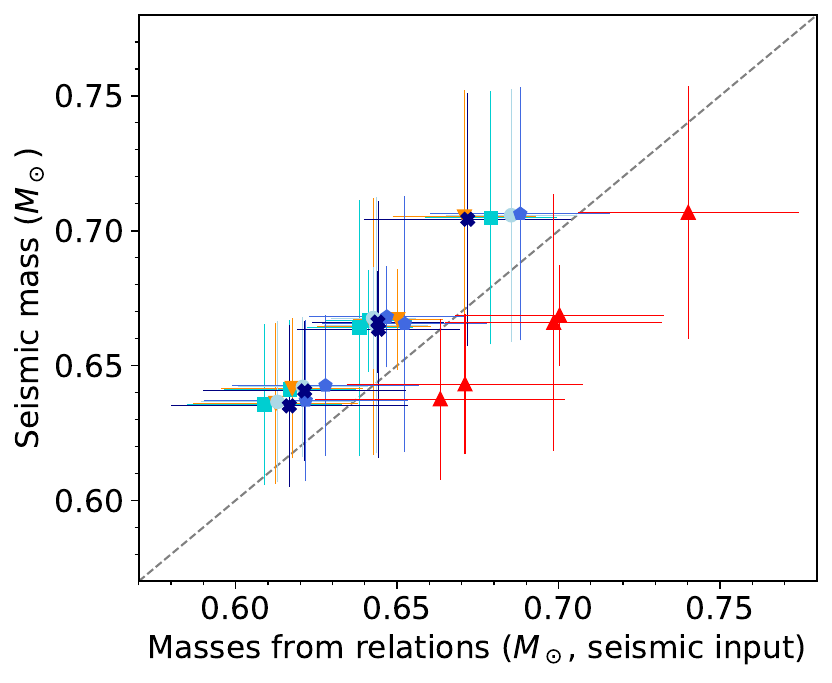}
   \caption{Comparison of the RRc masses with results from various mass relations based on physical parameters. On the left: masses calculated from classical observational constraints obtained for $L$ and \teff{}. Right: masses calculated from the $L$ and \teff{} outputs produced by our seismic results listed in Table~\ref{tab:output_values}. Seismic masses are shifted by small amounts in both plots to make individual error bars visible. The dashed line is the identity line in both plots. }
              \label{fig:mass_comp}%
    \end{figure*}

\subsection{Comparison of the rHB and RRc masses}

As we discussed in Section~\ref{sect:intro}, estimating the masses of RR Lyrae stars in a model-independent way is an extremely difficult task. While evolutionary models can be used to constrain HB masses in globular clusters \citep[see, e.g.,][]{Gratton-2010,Howell-2024,kumar-2024}, and masses for RRd stars can be estimated from non-linear models \citep{Molnar-2015}, these approaches still need independent verification. Here we present, for the first time, simultaneous mass estimates based on two independent seismic approaches for stars along the HB in a globular cluster.
And although these seismic models and scaling relations could still contain their own model uncertainties, they can now be verified against each other. 

For M4, we found a combined HB mass of $0.652 \pm 0.022$~M$_\odot$. If we split this sample into their constituent RR Lyrae and rHB parts, we find average masses of $M_{\rm rHB} = 0.657 \pm 0.034$~M$_\odot$ and $M_{\rm RRc} = 0.648 \pm 0.028$~M$_\odot$, respectively. The small difference of $\approx0.01$~M$_\odot$ is in agreement with the evolutionary predictions that stellar (envelope) masses along the HB decrease towards the blue \citep{salaris-2006}. Therefore, we can conclude that our two seismic mass estimates, the scaling relations for rHB stars, and the linear RRc model method of \citet{netzel2022}, are in agreement. Thus, we may conclude that fitting the $f_{61}$ modes offers a reliable way to estimate RR Lyrae star masses, even if it is only available for overtone stars.

\subsection{Comparison with mass relations and models}
Relations to calculate masses for RR Lyrae, or more broadly for HB stars from observational constraints, have been given by multiple authors, derived from pulsation and evolutionary models. Here we compare seven prescriptions with our seismic masses. The first widely adopted relation on dependence of the pulsation period on the physical parameters (\textit{M, L,} \teff{}) was published by \citet{vAB1971}. This was later refined and updated by multiple authors, also incorporating a metallicity term \citep{Bono-1997,jurcsik-1998,Marconi-2003,diCri-2004,Marconi-2015}. A more general relation between mass, color and metallicity was given for HB stars by \citet{Gratton-2010}.

The accuracy of these prescriptions, however, strongly depend on the  modeling choices and uncertainties intrinsic to parameter choices, which are not often explored. Furthermore, many prescriptions are based on \teff{} and \textit{L} values, which have to be converted from observations and have many uncertainties and potential parameter degeneracies themselves. Converting apparent brightness to theoretical luminosity, for example, involves uncertainties in the distance of the object, in the amount of interstellar absorption between the observer and the object, and even in the calculation of the bolometric correction. Similarly, converting observed colors to \teff{} involves uncertainties coming from interstellar absorption and the choice of extinction law parameter, as well as from the accuracy of the calibration of the color--\teff{} conversion scales themselves. 

    \begin{figure}
   \centering
   \includegraphics[width=1.0\columnwidth]{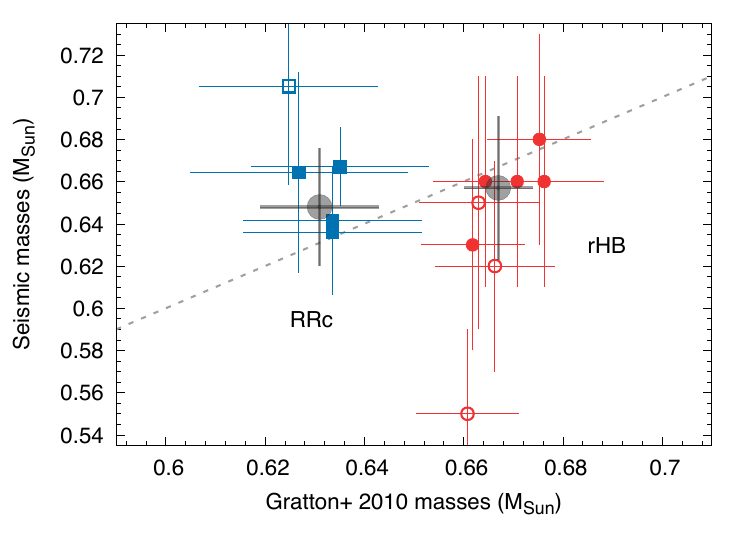}
   \caption{Comparison of the seismic HB masses, both for the RRc and rHB stars, with the mass relation published by \citet{Gratton-2010}. Empty symbols indicate marginal detections. Large grey symbols indicate the averages for both groups. The dashed line is the identity line.  }
              \label{fig:mass_comp_gr}%
    \end{figure}

    \begin{figure}
   \centering
   \includegraphics[width=1.0\columnwidth]{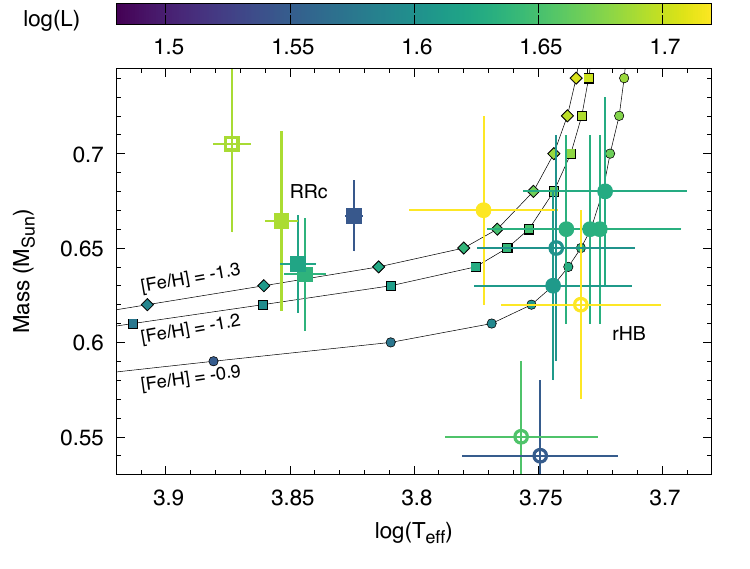}
   \caption{Comparison of the seismic HB masses to BaSTI ZAHB model sequences. All three models are $\alpha$-enhanced ([$\alpha$/Fe] = 0.4), and have similar He abundances, but differing metal contents. Colors indicate luminosities; squares represent RRc stars whereas circles represent RHB stars; empty symbols indicate marginal detections.  }
              \label{fig:basti}%
    \end{figure}

With these limitations in mind, we calculated the RR Lyrae masses using these relations and compared them to our seismic results in two ways. First we used the classical constraints, $L$ and \teff{} values, computed from the Johnson photometry, which also served as initial values for our seismic model fits. These results are plotted in the left panel of Fig.~\ref{fig:mass_comp}. As the figure shows, the values are generally in the right range, between 0.45 and 1.0~$M_\odot$, however, differences from the seismic values can reach 0.1--0.3~$M_\odot$.

We then recalculated the masses using the $L$ and \teff{} values obtained from the seismic model fits listed in Table~\ref{tab:output_values}, and plotted them in the right panel of Fig.~\ref{fig:mass_comp}. In contrast with the results above, these correlate very well with our seismic masses. The matches are not perfect, however, as clear systematic differences are visible. Most relations give results very similar to the original one published by \citet{vAB1971}, underestimating the masses by an average difference of --0.020 to --0.028~$M_\odot$, relative to the seismic masses. The only relation that overestimates the masses is that of \citet{Marconi-2015}, with an average difference of +0.032~$M_\odot$. These results highlight the true difference between various models and their parameter choices, rather than the uncertainties coming from the observations. The main source of observational uncertainty left in these fits is in the frequencies of the modes we fit.

\citet{Gratton-2010} determined a different mass relation based on the $B-V$ (or $V-I$) color indices and [Fe/H] indices of HB stars. We calculated the masses of both the RRc and the rHB stars using our dereddened $B-V$ colors. The results are displayed in Fig.~\ref{fig:mass_comp_gr}. This relation results in a larger mass difference (0.037~$M_\odot$) between the two groups than what we found from the seismic data. This difference could be related to he fact that the relation is based on zero-age HB stellar models, whereas the stars in M4 are not necessarily on the zero-age horizontal branch (ZAHB), but rather at evolutionary stages dictated by their common ages and thus common masses. 

Finally, we compared our results directly to the BaSTI ZAHB models. We plot three model sequences in Fig.~\ref{fig:basti} on the $\log{}(T_{\rm eff})$ -- mass plane. All three models are $\alpha$-enhanced ([$\alpha$/Fe] = +0.4), with differing [Fe/H] indices (-1.3, -1.2, -0.9) but similar He contents (Y = 0.259, 0.250, 0.252). While the inferred mass and \teff{} values do not line up well with a single track, three out of four RRc stars within $1\sigma$ from the measured [Fe/H] = $-1.1\pm 0.07$ value of the cluster, with RRc stars aligning with slightly more metal-poor (more massive), and RHB stars with slighly metal-rich (lighter) sequences. This result agrees with what we observe in Fig.~\ref{fig:mass_comp_gr}. However, whether the ZAHB models overestimate the mass gradient along the HB or our mass measurements contain systematic offsets cannot be decided from these results alone.

\section{Conclusions}
\label{sect:concl}
Determining the masses of RR Lyrae variables is a notoriously difficult task. While relations based on physical parameters exist, and pulsational masses can, in principle, provide masses for double-mode stars, these are strongly model-dependent methods that could not be verified by independent mass estimates. Asteroseismology offers a new way to determine stellar masses \citep{aerts-2021}. While seismic masses still depend on stellar envelope models and on the validity of scaling relations, they can be tested against other observational techniques, e.g., dynamical masses in eclipsing binaries. 

Globular clusters, such as M4, provide an opportunity to probe the late stages of low-mass stellar evolution. Building on the unique capabilities of the \textit{Kepler} space telescope, we calculated new, more accurate masses for the oscillating giant stars using asteroseismic scaling relations and $\nu_{\rm max}$ values determined with the new \texttt{pyMON} code (Howell et al., in prep.). We then analysed the overtone RR Lyrae stars in the cluster and determined their masses via peak-bagging and fitting the frequencies with seismic model calculations \citet{netzel2022,Netzel-2023}. With these, we were able to relate RR~Lyrae seismic masses to seismic masses of red HB stars, and found that they match each other closely, with the RR~Lyrae models being lighter, as expected from stellar evolution. We found the average rHB and RRc masses to be $M_{\rm rHB} = 0.657 \pm 0.034$~M$_\odot$ and $M_{\rm RRc} = 0.648 \pm 0.028$~M$_\odot$, respectively.

Furthermore, we estimate a low He enrichment in the cluster, in agreement with the findings of \citet{Villanova-2012}, based on the RRc model results. However, the models leave the chemical compositions ambiguous for multiple targets, which makes this result uncertain.

While our result is not yet a direct measurement of RR Lyrae masses, as it relies on the assumption that RR Lyrae and rHB stars in the cluster have very similar masses, we were able to test model-dependent seismic (or pulsational) masses against independent mass estimates for the first time. We find that while the RRc stars have slightly lower masses, as expected from the stucture of the HB, the difference of $\approx 0.01$~M$_\odot$ is less than what stellar evolutionary models predict. {Whether the source of this discrepancy comes from the pulsational or the evolutionary models remains to be seen: but even if considering such systematic differences, we estimate that our method to determine RR~Lyrae masses is accurate to $\sigma_M \approx0.05$\,M$_\odot$.

We also compared the modeled physical parameters of the RRc stars to various relations between \textit{M, L, \teff{}} and the pulsation periods, and found that while these generally agree, systematical differences do exist. We also show, however, that the usability of these relations is rather limited, as \textit{L} and \teff{} values have to be determined very accurately, otherwise the calculated masses will be very uncertain. 

This study not only tests RR Lyrae masses for the first time, but it also lends further support to the hypothesis of the $f_{61}$ modes being high-order $\ell$ modes, as proposed by  \citet{dziem-2016}. Therefore, it is now possible to estimate masses (and $\log g$ values) for field RRc and RRd stars to within a few hundredth M$_\odot$ accuracy, if enough modes can be detected from high-precision photometry \citep{netzel2022,Netzel-2023}. Furthermore, RRc stars now offer a way to measure masses on the HB in globular clusters, even where rHB stars are too faint for asteroseismology. RR Lyrae stars thus open up a new way to study mass loss in more clusters, either with \textit{Kepler} (Kalup et al., in prep.), or with future instruments, such as the \textit{Roman} Space Telescope \citep{Molnar-WP-2023} or the HAYDN telescope project, which is designed specifically to observe dense stellar fields \citep{miglio-2021}.

\begin{acknowledgements}
This research was supported by the `SeismoLab' KKP-137523 \'Elvonal grant and the NKFIH excellence grant TKP2021-NKTA-64 of the Hungarian Research, Development and Innovation Office (NKFIH), and by the LP2025-14/2025
Lendület grant of the Hungarian Academy of Sciences. Cs.K.\ was supported by the \'UNKP-23-3, New National Excellence Program of the Ministry of Culture and Innovation from the source of the National Research, Development and Innovation Fund. This research has received support from the European Research Council (ERC) under the European Union’s Horizon 2020 research and innovation programme (Grant Agreement No. 947660). H.N.\ is funded by the Swiss National Science Foundation (award PCEFP2\_194638) and the European Research Council (ERC) under the European Union’s Horizon 2020 research and innovation program (grant agreement No. 951549 - UniverScale). M.J.\ gratefully acknowledges funding of MATISSE: \textit{Measuring Ages Through Isochrones, Seismology, and Stellar Evolution}, awarded through the European Commission's Widening Fellowship. This project has received funding from the European Union's Horizon 2020 research and innovation programme. This research made use of NASA’s Astrophysics Data System Bibliographic Services, as well as of the SIMBAD database operated at CDS, Strasbourg, France.
\end{acknowledgements}

\bibliographystyle{aa} 
\bibliography{m4masses}

\appendix
\section{V61, the modulated star}
\label{sec:v61}
The light curve of V61 has a much lower pulsation amplitude compared to the others and shows amplitude and phase modulations. This pattern is very similar to the modulated low-amplitude RRc stars observed in M80 \citet{Molnar-2023} and elsewhere \citep{antipin-2005IBVS,smolec-2017,netzel2019}. Just like in those cases, we detect asymmetry side peaks next to the pulsation frequency, in this case, above it. We cut the light curve into 1.5~d long segments and fitted $A_1$ and $\phi_1$ for each segment separately with the same, fixed $f_1$ frequency to study the shape of the modulation more closely. 

As Figure~\ref{fig:v61-bl} shows, the modulation pattern is multiperiodic. We calculated the Fourier-spectra of both the amplitude and phase variations. The former is also shown in Fig.~\ref{fig:v61-bl}; the spectrum of the phase variation looks very similar. Both show to separate frequency peak, from which we calculated two modulation periods: $P_{m1} = 11.39\pm 0.05$~d and $P_{m2} = 19.63\pm 0.25$~d.

\section{Data tables}
\label{sect:appx}
In Table \ref{tab:lc}, we present the reduced K2 light curves for the five RRc stars analysed in this work. The full table will be available online. In Table \ref{tab:numax}, we list the updated seismic and physical parameters for the RGB, AGB and rHB stars. In Tables~\ref{tab:grat} and \ref{tab:mass_comp} we list the mass estimates we show in Figures~\ref{fig:mass_comp} and \ref{fig:mass_comp_gr}.

\section{Corner plots}
\label{sect:corner}
The corner plots for each RRc star are shown in Figs.~\ref{fig:v06c}, \ref{fig:v42c}, \ref{fig:v43c}, \ref{fig:v61c} and \ref{fig:v76c}, respectively. The diagonal panels display histograms of the parameter distributions, with vertical lines indicating the mean values and their associated uncertainties, which are also annotated at the top of each column. The peak of the mass distribution generally coincides with its mean value. Interestingly, the distribution of the hydrogen content $X$ shows bimodality, where two distinct maxima appear at the boundaries of the accepted range. This effect is most pronounced for V06 and V42.

Several parameter correlations are evident. As expected, there is a clear positive correlation between effective temperature and luminosity, which follows the lines of constant period. Additionally, for V06 and V42, an anticorrelation  between effective temperature and mass, as well as between luminosity and mass is visible. This is also expected as mass and effective temperature are anticorrelated along the horizontal branch.

        \begin{figure}
   \centering
   \includegraphics[width=1.0\columnwidth]{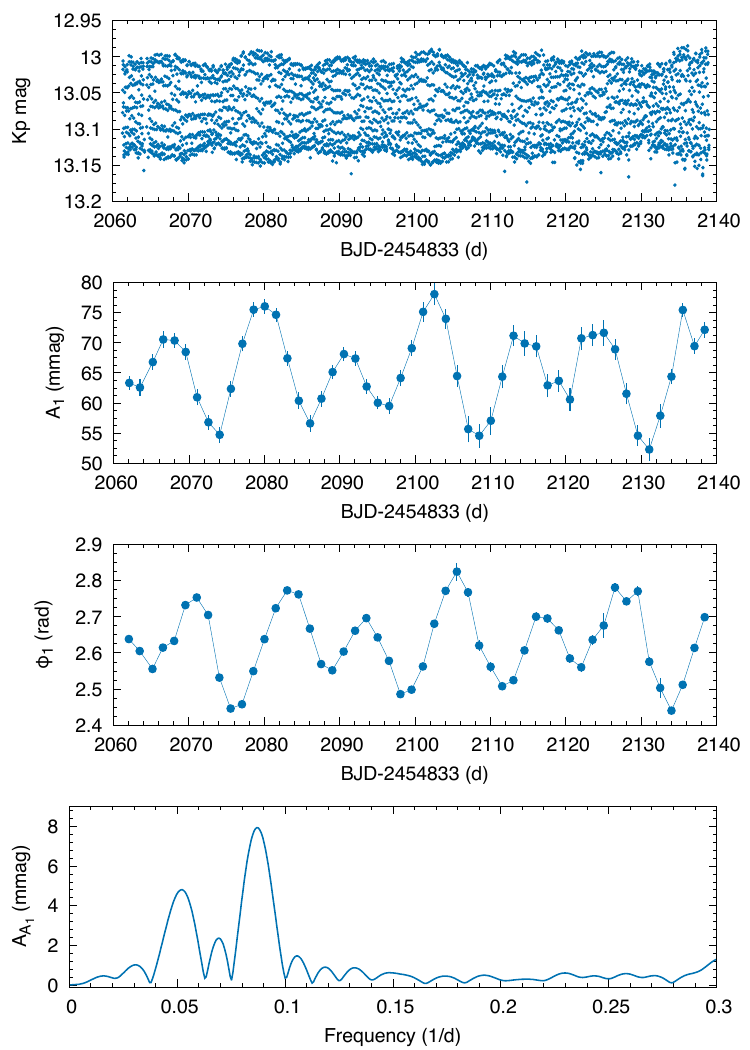}
   \caption{V61 amplitude and phase modulation}
              \label{fig:v61-bl}%
    \end{figure}

\begin{figure}
   \centering
   \includegraphics[width=1.0\columnwidth]{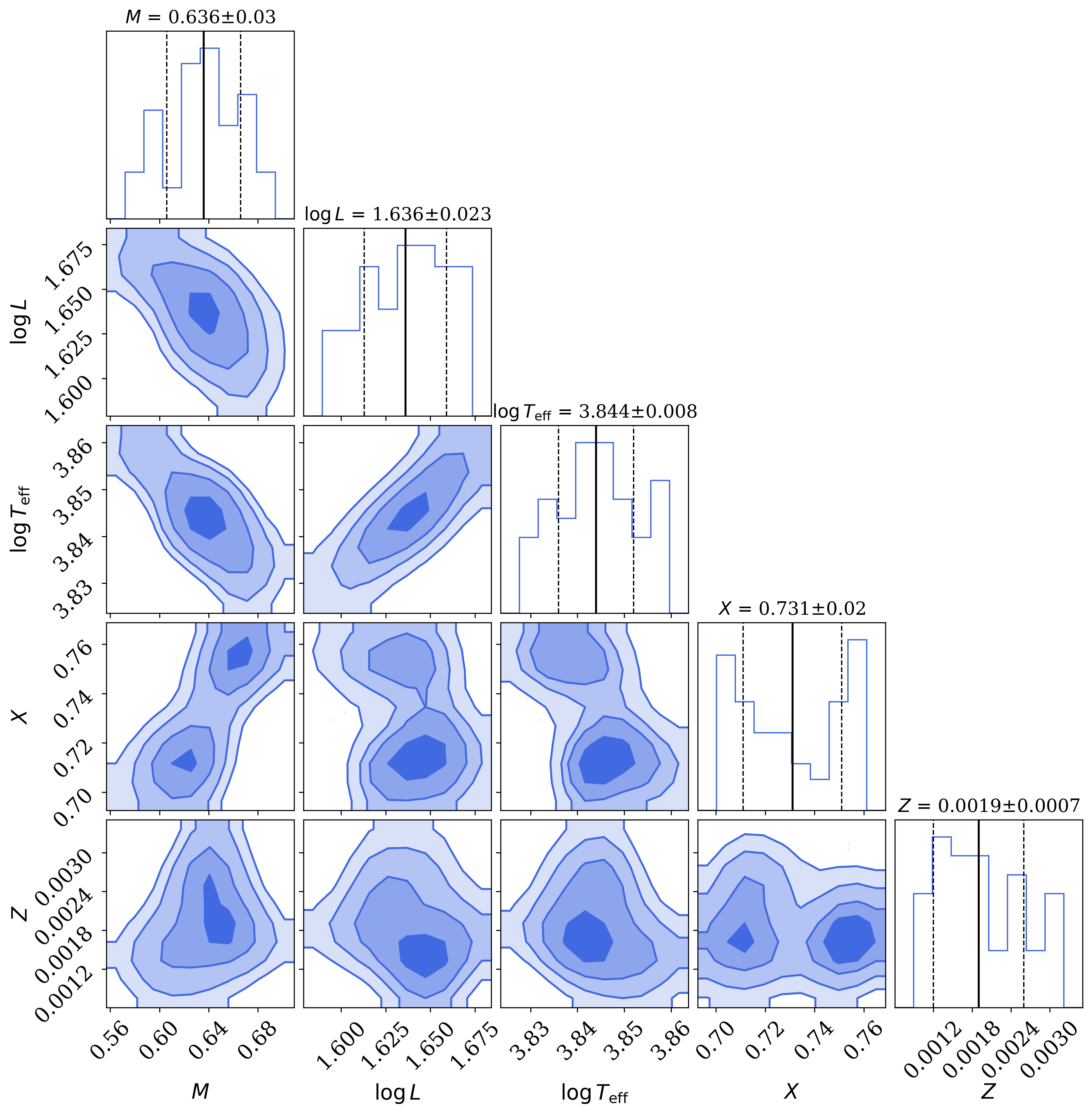}
   \caption{Corner plot for V06. Solid and dashed black lines in the diagonal panels correspond to the mean values and their standard deviation as indicated on top of each column.}
              \label{fig:v06c}%
    \end{figure}

\begin{figure}
   \centering
   \includegraphics[width=1.0\columnwidth]{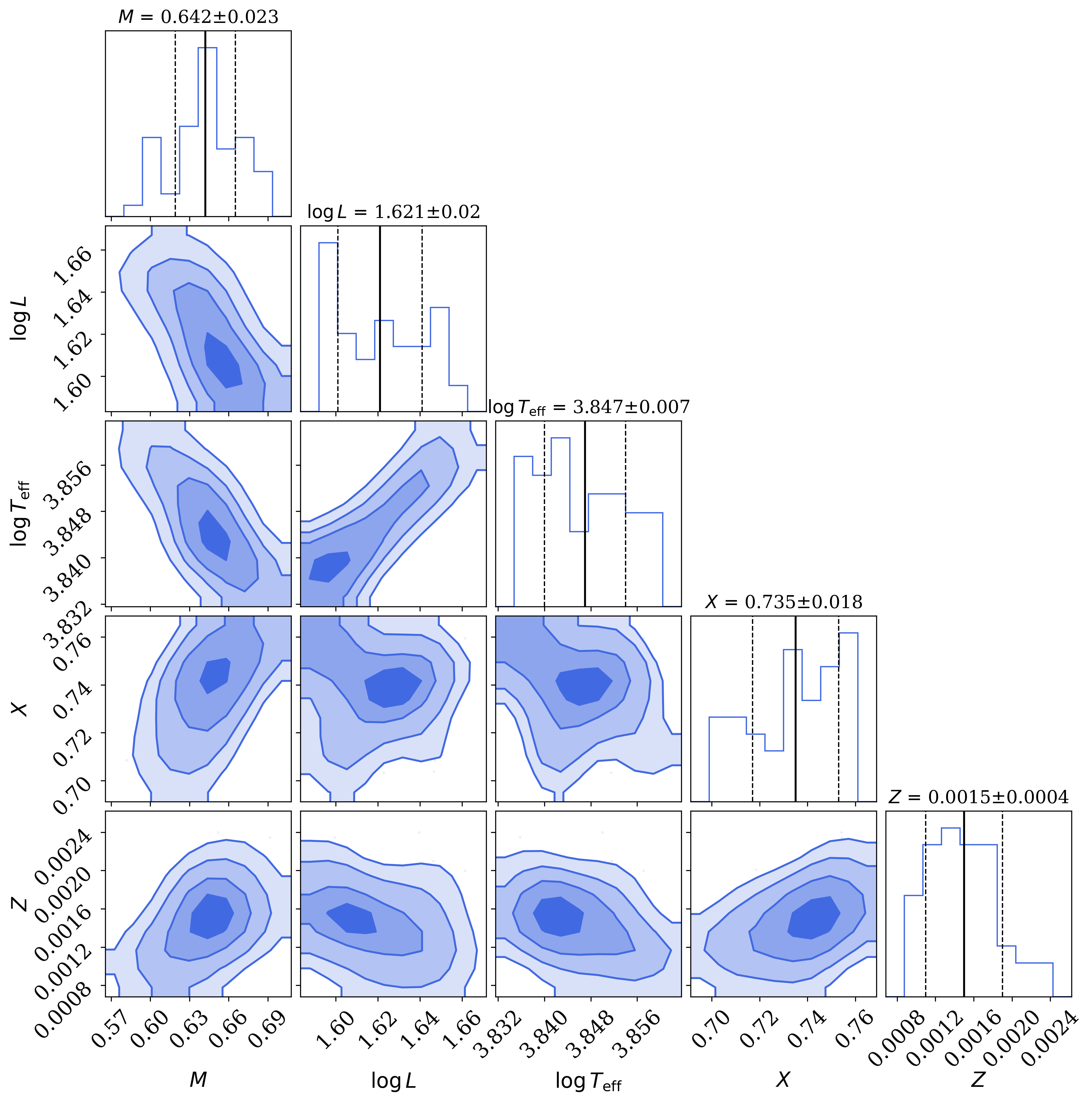}
   \caption{Same as Fig.~\ref{fig:v06c}] but for V42.}
              \label{fig:v42c}%
    \end{figure}

\begin{figure}
   \centering
   \includegraphics[width=1.0\columnwidth]{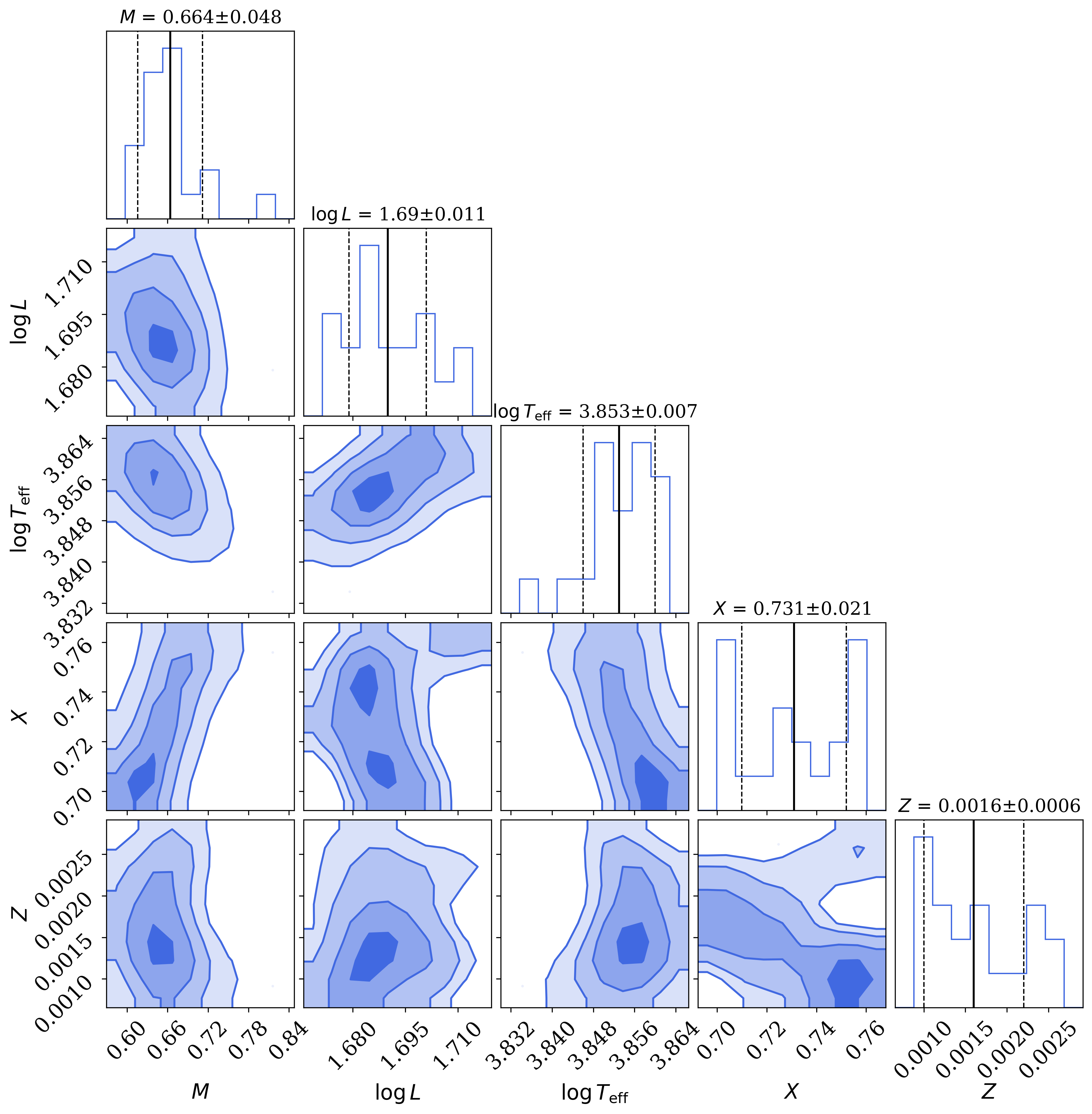}
   \caption{Same as Fig.~\ref{fig:v06c}] but for V43.}
              \label{fig:v43c}%
    \end{figure}

\begin{figure}
   \centering
   \includegraphics[width=1.0\columnwidth]{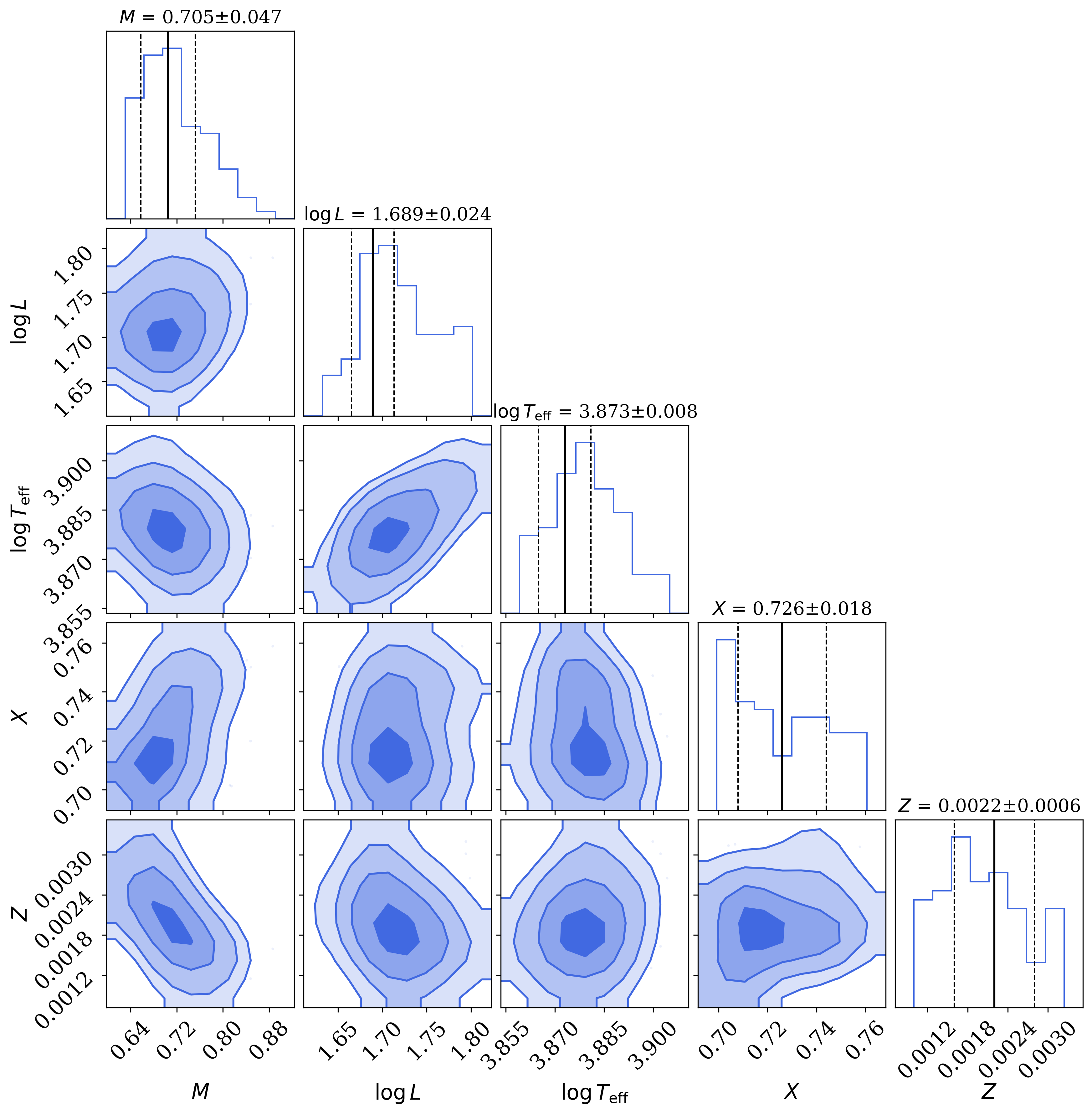}
   \caption{Same as Fig.~\ref{fig:v06c}] but for V61.}
              \label{fig:v61c}%
    \end{figure}

\begin{figure}
   \centering
   \includegraphics[width=1.0\columnwidth]{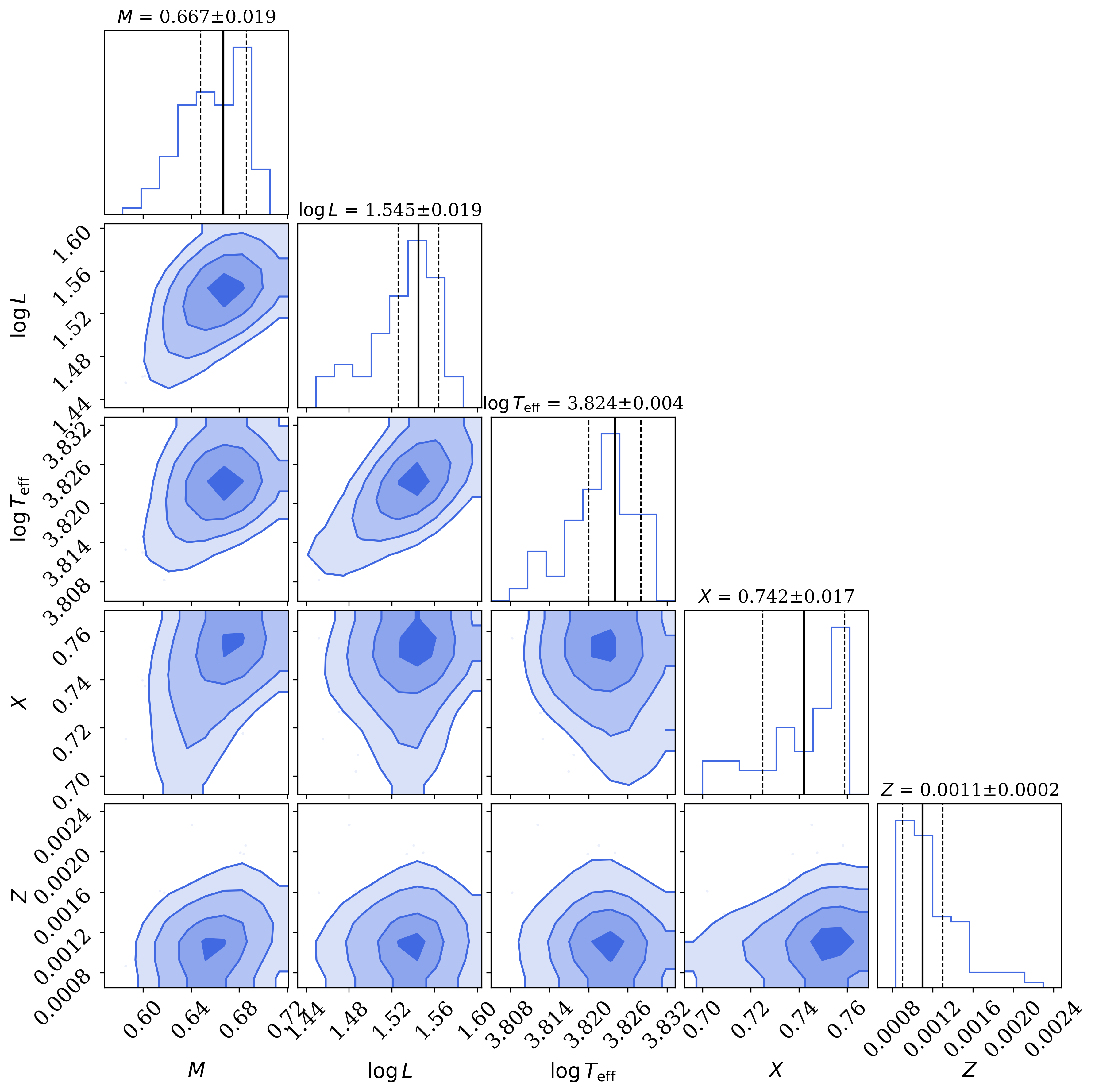}
   \caption{Same as Fig.~\ref{fig:v06c}] but for V76.}
              \label{fig:v76c}%
    \end{figure}

\begin{table}[!h]
    
    \caption{K2SC- and trend-corrected photometric data for the five RRc stars. The full table is available in machine-readable format online.}
    \label{tab:lc}
    \centering
    \begin{tabular}{l c c c}
\hline
\hline
ID & BJD--2454833 & Kp    & $\Delta $Kp \\
~  &    (d)       & (mag) &   (mag     \\
\hline
V06 &  2060.284202 & 13.41794 & 0.00102 \\
V06 &  2060.304635 & 13.34582 & 0.00094 \\
V06 &  2060.325067 & 13.22527 & 0.00084 \\
V06 &  2060.345499 & 13.16045 & 0.00080 \\
V06 &  2060.365931 & 13.12550 & 0.00078 \\
V06 &  2060.386363 & 13.09217 & 0.00076 \\
V06 &  2060.406796 & 13.08390 & 0.00076 \\
V06 &  2060.427228 & 13.10122 & 0.00077 \\
\multicolumn{4}{l}{\dots}\\
\hline
\end{tabular}
\end{table}

\begin{table}
    
    \caption{Mass values for RR Lyrae and rHB stars based on the mass relation of \citet{Gratton-2010}, compared to our seismic masses.}
    \label{tab:grat}
    \centering
    \begin{tabular}{l c c}
\hline
\hline
ID & $M_{\rm Gratton2010}$ & $M_{\rm seism}$ \\
V06 & $0.634\pm0.018$ & $0.636\pm0.030$ \\
V42 & $0.634\pm0.018$ & $0.642\pm0.026$ \\
V43 & $0.627\pm0.022$ & $0.664\pm0.048$ \\
V76 & $0.635\pm0.018$ & $0.667\pm0.019$ \\
V61 & $0.627\pm0.018$ & $0.705\pm0.047$ \\
M4RHB39 & $0.671\pm0.010$ & $0.66\pm0.05$ \\
M4RHB41 & $0.675\pm0.011$ & $0.68\pm0.05$ \\
M4RHB42 & $0.662\pm0.010$ & $0.63\pm0.05$ \\
M4RHB44 & $0.664\pm0.010$ & $0.66\pm0.05$ \\
M4RHB127 & $0.661\pm0.010$ & $0.55\pm0.04$ \\
M4RHB196 & $0.676\pm0.012$ & $0.66\pm0.05$ \\
M4RHB201 & $0.666\pm0.012$ & $0.62\pm0.05$ \\
M4RHB202 & $0.663\pm0.012$ & $0.65\pm0.06$ \\
\hline
\end{tabular}
\end{table}

\onecolumn

\begin{table*}
    
    \caption{Mass values calculated from the various mass relations, using either the classical observational constraints or the physical parameters from our seismic fits as constraints.}
    \label{tab:mass_comp}
    \centering
    \begin{tabular}{lccccccc}
\hline
\hline
ID &  $M_{\rm seism}$ &   vA-B1971      & B1997           &  J1998          &  M2003          & dC2004          & M2015 \\
\hline
\multicolumn{8}{c}{classical constraints}\\
\hline
V06 & $0.636\pm0.030$ & $0.684\pm0.089$ & $0.681\pm0.057$ & $0.696\pm0.061$ & $0.684\pm0.057$ & $0.690\pm0.061$ & $0.750\pm0.060$ \\
V42 & $0.642\pm0.026$ & $0.740\pm0.082$ & $0.746\pm0.052$ & $0.764\pm0.055$ & $0.748\pm0.052$ & $0.750\pm0.056$ & $0.826\pm0.055$ \\
V43 & $0.664\pm0.048$ & $0.773\pm0.152$ & $0.778\pm0.097$ & $0.800\pm0.102$ & $0.783\pm0.096$ & $0.787\pm0.099$ & $0.867\pm0.093$ \\
V76 & $0.667\pm0.019$ & $0.882\pm0.081$ & $0.902\pm0.050$ & $0.954\pm0.052$ & $0.897\pm0.051$ & $0.887\pm0.054$ & $1.016\pm0.051$ \\
(V61) & $0.705\pm0.047$ & $0.466\pm0.177$ & $0.456\pm0.117$ & $0.430\pm0.133$ & $0.464\pm0.114$ & $0.475\pm0.117$ & $0.480\pm0.122$ \\
\hline 
\multicolumn{8}{c}{seismic constraints}\\
\hline
V06 & $0.636\pm0.030$ & $0.617\pm0.037$ & $0.609\pm0.024$ & $0.612\pm0.025$ & $0.613\pm0.024$ & $0.622\pm0.032$ & $0.663\pm0.039$ \\
V42 & $0.642\pm0.026$ & $0.621\pm0.031$ & $0.617\pm0.020$ & $0.617\pm0.022$ & $0.621\pm0.020$ & $0.628\pm0.029$ & $0.671\pm0.037$ \\
V43 & $0.664\pm0.048$ & $0.644\pm0.025$ & $0.638\pm0.016$ & $0.643\pm0.018$ & $0.644\pm0.016$ & $0.652\pm0.026$ & $0.699\pm0.034$ \\
V76 & $0.667\pm0.019$ & $0.644\pm0.020$ & $0.641\pm0.013$ & $0.650\pm0.014$ & $0.643\pm0.013$ & $0.647\pm0.024$ & $0.700\pm0.032$ \\
(V61) & $0.705\pm0.047$ & $0.672\pm0.032$ & $0.679\pm0.020$ & $0.671\pm0.022$ & $0.685\pm0.020$ & $0.688\pm0.028$ & $0.740\pm0.034$ \\
\hline
\end{tabular}
\end{table*}

\begin{longtable}{lcccccccccccc}
\caption{\label{tab:numax} The updated global seismic properties and calculated physical parameters for the giants in M4. }\\
\hline\hline
ID & Type &  QF & \textit{G} & $\nu_{\rm max}$ & $\Delta \nu_{\rm max}$  & \teff{} & $\Delta$\teff{} & \textit{L} & $\Delta L_{\rm rand}$ & $\Delta L_{\rm sys}$ & \textit{M} & $\Delta M$ \\
~ & ~ & ~ & (mag) & ($\mu$Hz) & ($\mu$Hz) & (K) & (K) & ($L_\odot$) & ($L_\odot$) & ($L_\odot$)  & ($M_\odot$) & ($M_\odot$) \\ 
\hline
\endfirsthead
\caption{continued.}\\
\hline\hline
ID & Type &  Quality & \textit{G} & $\nu_{\rm max}$ & $\Delta \nu_{\rm max}$  & \teff{} & $\Delta$\teff{} & \textit{L} & $\Delta L_{\rm rand}$ & $\Delta L_{\rm sys}$ & \textit{M} & $\Delta M$ \\
~ & ~ & ~ & (mag) & ($\mu$Hz) & ($\mu$Hz) & (K) & (K) & ($L_\odot$) & ($L_\odot$) & ($L_\odot$)  & ($M_\odot$) & ($M_\odot$) \\ 
\hline
\endhead
\hline
\endfoot
M4AGB01 & AGB & MD & 12.08 & 9.60 & 0.40 & 4944 & 29 & 104.1 & 0.9 & 12.1 & 0.56 & 0.03 \\
M4AGB02 & AGB & D & 11.84 & 6.67 & 0.13 & 4838 & 62 & 128.3 & 2.6 & 14.9 & 0.51 & 0.03 \\
M4AGB59 & AGB & MD & 11.36 & 4.73 & 0.18 & 5029 & 108 & 209.0 & 5.5 & 24.3 & 0.52 & 0.04 \\
M4AGB62 & AGB & D & 11.93 & 8.08 & 0.09 & 4941 & 52 & 125.3 & 1.9 & 14.6 & 0.56 & 0.02 \\
M4RGB10 & RGB & D & 13.40 & 48.50 & 0.47 & 4960 & 65 & 32.3 & 0.6 & 3.7 & 0.86 & 0.04 \\
M4RGB11 & RGB & D & 12.22 & 10.74 & 0.22 & 4612 & 58 & 92.9 & 2.4 & 10.8 & 0.71 & 0.04 \\
M4RGB13 & RGB & D & 13.53 & 49.20 & 1.99 & 4840 & 108 & 27.0 & 0.9 & 3.1 & 0.79 & 0.07 \\
M4RGB14 & RGB & D & 13.09 & 33.41 & 0.19 & 4807 & 59 & 41.2 & 0.8 & 4.8 & 0.84 & 0.04 \\
M4RGB15 & RGB & MD & 12.59 & 17.55 & 0.76 & 4865 & 51 & 65.2 & 1.1 & 7.6 & 0.67 & 0.04 \\
M4RGB16 & RGB & D & 13.30 & 44.02 & 0.76 & 4921 & 62 & 30.8 & 0.6 & 3.6 & 0.77 & 0.04 \\
M4RGB17 & RGB & D & 13.55 & 51.89 & 0.71 & 4785 & 61 & 28.4 & 0.6 & 3.3 & 0.92 & 0.05 \\
M4RGB20 & RGB & MD & 12.10 & 10.86 & 0.35 & 4535 & 59 & 103.0 & 3.0 & 11.9 & 0.84 & 0.05 \\
M4RGB21 & RGB & MD & 12.93 & 31.62 & 0.95 & 4855 & 59 & 44.2 & 0.8 & 5.1 & 0.83 & 0.05 \\
M4RGB22 & RGB & D & 13.23 & 36.93 & 1.42 & 4852 & 61 & 35.4 & 0.7 & 4.1 & 0.78 & 0.05 \\
M4RGB23 & RGB & D & 13.38 & 42.55 & 1.04 & 4897 & 108 & 31.3 & 1.0 & 3.6 & 0.77 & 0.06 \\
M4RGB24 & RGB & D & 13.31 & 45.05 & 0.47 & 4829 & 108 & 31.8 & 1.1 & 3.7 & 0.86 & 0.07 \\
M4RGB25 & RGB & D & 12.70 & 19.79 & 0.57 & 4719 & 108 & 60.1 & 2.4 & 7.0 & 0.78 & 0.07 \\
M4RGB26 & RGB & D & 11.65 & 5.81 & 0.13 & 4541 & 79 & 158.5 & 6.1 & 18.4 & 0.69 & 0.05 \\
M4RGB27 & RGB & D & 13.78 & 69.85 & 0.85 & 4954 & 108 & 21.0 & 0.6 & 2.4 & 0.81 & 0.06 \\
M4RGB28 & RGB & MD & 14.39 & 126.69 & 2.51 & 5157 & 108 & 13.0 & 0.3 & 1.5 & 0.79 & 0.06 \\
M4RGB29 & RGB & D & 13.51 & 58.99 & 0.57 & 4986 & 60 & 24.7 & 0.4 & 2.9 & 0.79 & 0.04 \\
M4RGB30 & RGB & MD & 13.01 & 30.41 & 0.95 & 4913 & 67 & 43.1 & 0.8 & 5.0 & 0.75 & 0.04 \\
M4RGB31 & RGB & MD & 13.60 & 57.77 & 1.14 & 4966 & 63 & 25.9 & 0.5 & 3.0 & 0.82 & 0.04 \\
M4RGB32 & RGB & MD & 13.85 & 74.71 & 1.04 & 5026 & 65 & 20.4 & 0.3 & 2.4 & 0.80 & 0.04 \\
M4RGB34 & RGB & D & 14.28 & 116.59 & 2.46 & 5059 & 108 & 13.4 & 0.3 & 1.6 & 0.80 & 0.06 \\
M4RGB35 & RGB & D & 13.06 & 32.07 & 0.57 & 4838 & 77 & 42.1 & 1.0 & 4.9 & 0.81 & 0.05 \\
M4RGB36 & RGB & D & 13.03 & 34.24 & 1.42 & 4808 & 60 & 47.7 & 1.0 & 5.5 & 1.00 & 0.06 \\
M4RGB37 & RGB & D & 13.15 & 35.14 & 0.66 & 4821 & 108 & 39.6 & 1.4 & 4.6 & 0.85 & 0.07 \\
M4RGB48 & RGB & D & 12.56 & 17.88 & 0.22 & 4746 & 57 & 71.3 & 1.5 & 8.3 & 0.82 & 0.04 \\
M4RGB58 & RGB & D & 11.97 & 10.32 & 0.27 & 4840 & 108 & 122.3 & 4.1 & 14.2 & 0.76 & 0.06 \\
M4RGB104 & RGB & D & 13.15 & 33.73 & 0.57 & 4851 & 108 & 38.9 & 1.3 & 4.5 & 0.78 & 0.06 \\
M4RGB124 & RGB & D & 14.05 & 91.66 & 1.04 & 5088 & 108 & 17.1 & 0.4 & 2.0 & 0.79 & 0.06 \\
M4RGB129 & RGB & D & 13.77 & 70.17 & 1.99 & 4989 & 108 & 22.6 & 0.6 & 2.6 & 0.86 & 0.07 \\
M4RGB130 & RGB & D & 13.08 & 36.93 & 0.57 & 4897 & 108 & 40.2 & 1.2 & 4.7 & 0.85 & 0.07 \\
M4RGB131 & RGB & D & 14.07 & 96.96 & 1.61 & 5047 & 108 & 16.9 & 0.4 & 2.0 & 0.85 & 0.07 \\
M4RGB133 & RGB & D & 11.15 & 3.23 & 0.04 & 4377 & 108 & 260.0 & 17.1 & 30.1 & 0.72 & 0.08 \\
M4RGB142 & RGB & D & 13.86 & 74.97 & 1.56 & 4988 & 74 & 22.5 & 0.4 & 2.6 & 0.91 & 0.05 \\
M4RGB143 & RGB & D & 13.64 & 66.72 & 1.71 & 4937 & 108 & 23.2 & 0.7 & 2.7 & 0.87 & 0.07 \\
M4RGB147 & RGB & D & 14.02 & 103.68 & 1.61 & 5218 & 108 & 18.5 & 0.4 & 2.1 & 0.88 & 0.07 \\
M4RGB148 & RGB & D & 12.45 & 12.29 & 0.53 & 4632 & 54 & 76.4 & 1.8 & 8.9 & 0.66 & 0.04 \\
M4RGB165 & RGB & D & 13.99 & 91.85 & 2.09 & 5287 & 108 & 23.0 & 0.4 & 2.7 & 0.93 & 0.07 \\
M4RGB169 & RGB & D & 13.75 & 80.08 & 2.04 & 5143 & 73 & 24.8 & 0.4 & 2.9 & 0.96 & 0.05 \\
M4RGB180 & RGB & MD & 12.90 & 23.88 & 0.57 & 4972 & 66 & 54.0 & 1.0 & 6.3 & 0.70 & 0.04 \\
M4RGB188 & RGB & D & 13.05 & 30.98 & 0.57 & 4810 & 108 & 44.0 & 1.5 & 5.1 & 0.84 & 0.07 \\
M4RGB190 & RGB & MD & 13.67 & 61.41 & 1.61 & 4949 & 108 & 23.1 & 0.7 & 2.7 & 0.79 & 0.06 \\
M4RGB191 & RGB & MD & 13.37 & 40.25 & 0.57 & 4892 & 61 & 32.9 & 0.6 & 3.8 & 0.76 & 0.04 \\
M4RGB194 & RGB & MD & 13.98 & 80.21 & 1.23 & 5035 & 72 & 18.0 & 0.3 & 2.1 & 0.76 & 0.04 \\
M4RGB195 & RGB & D & 12.08 & 8.65 & 0.09 & 4687 & 59 & 105.9 & 2.5 & 12.3 & 0.61 & 0.03 \\
M4RGB197 & RGB & D & 12.53 & 17.22 & 0.18 & 4734 & 108 & 72.7 & 2.8 & 8.5 & 0.81 & 0.07 \\
M4RGB205 & RGB & D & 13.70 & 62.18 & 0.76 & 4971 & 61 & 24.4 & 0.4 & 2.8 & 0.83 & 0.04 \\
M4RGB210 & RGB & D & 14.18 & 102.33 & 1.52 & 5093 & 46 & 15.2 & 0.2 & 1.8 & 0.78 & 0.03 \\
M4RGB213 & RGB & D & 13.67 & 53.68 & 0.85 & 4836 & 73 & 25.6 & 0.6 & 3.0 & 0.83 & 0.05 \\
M4RGB215 & RGB & D & 14.19 & 127.78 & 2.56 & 5174 & 92 & 15.0 & 0.6 & 1.7 & 0.91 & 0.07 \\
M4RGB216 & RGB & MD & 13.04 & 29.57 & 0.47 & 4813 & 108 & 46.3 & 1.6 & 5.4 & 0.84 & 0.07 \\
M4RGB217 & RGB & D & 13.36 & 51.25 & 2.18 & 5015 & 60 & 36.7 & 0.6 & 4.3 & 0.99 & 0.06 \\
M4RGB225 & RGB & D & 11.20 & 3.39 & 0.04 & 4427 & 55 & 258.5 & 8.5 & 30.0 & 0.72 & 0.04 \\
M4RGB229 & RGB & D & 12.51 & 15.95 & 0.18 & 4677 & 54 & 73.9 & 1.6 & 8.6 & 0.80 & 0.04 \\
M4RGB238 & RGB & MD & 13.91 & 83.79 & 1.71 & 5065 & 74 & 20.3 & 0.4 & 2.4 & 0.87 & 0.05 \\
M4RGB244 & RGB & D & 13.05 & 32.13 & 0.19 & 4877 & 62 & 44.2 & 0.8 & 5.1 & 0.83 & 0.04 \\
M4RGB252 & RGB & D & 13.58 & 55.28 & 0.66 & 5012 & 60 & 26.7 & 0.4 & 3.1 & 0.78 & 0.04 \\
M4RGB358 & RGB & MD & 12.94 & 27.02 & 1.14 & 4814 & 90 & 47.1 & 1.4 & 5.5 & 0.78 & 0.06 \\
M4RGB363 & RGB & MD & 11.90 & 7.93 & 0.22 & 4633 & 78 & 128.8 & 4.2 & 15.0 & 0.71 & 0.05 \\
M4RGB365 & RGB & MD & 11.71 & 6.41 & 0.14 & 4540 & 51 & 152.9 & 3.9 & 17.7 & 0.73 & 0.04 \\
M4RGB408 & RGB & MD & 12.44 & 13.61 & 0.66 & 4672 & 108 & 80.8 & 3.4 & 9.4 & 0.75 & 0.07 \\
M4RHB127 & RHB & MD & 12.99 & 36.61 & 1.33 & 5712 & 108 & 45.1 & 0.5 & 5.2 & 0.55 & 0.04 \\
M4RHB39 & RHB & D & 13.00 & 36.88 & 1.61 & 5358 & 108 & 42.8 & 0.7 & 5.0 & 0.66 & 0.05 \\
M4RHB41 & RHB & D & 13.04 & 36.42 & 0.57 & 5284 & 108 & 42.4 & 0.8 & 4.9 & 0.68 & 0.05 \\
M4RHB42 & RHB & D & 13.00 & 41.85 & 1.75 & 5545 & 108 & 40.7 & 0.6 & 4.7 & 0.63 & 0.05 \\
M4RHB44 & RHB & D & 13.03 & 39.70 & 1.33 & 5477 & 108 & 42.9 & 0.7 & 5.0 & 0.66 & 0.05 \\
M4RHB47 & RHB & MD & 13.01 & 29.45 & 1.61 & 5568 & 108 & 47.1 & 0.6 & 5.5 & 0.51 & 0.04 \\
M4RHB196 & RHB & D & 13.05 & 35.71 & 0.71 & 5310 & 108 & 42.9 & 0.8 & 5.0 & 0.66 & 0.05 \\
M4RHB201 & RHB & MD & 12.90 & 28.74 & 1.04 & 5404 & 108 & 52.6 & 0.9 & 6.1 & 0.62 & 0.05 \\
M4RHB202 & RHB & MD & 13.12 & 43.32 & 2.46 & 5529 & 108 & 40.1 & 0.6 & 4.7 & 0.65 & 0.06 \\
M4RHB246 & RHB & MD & 13.25 & 42.94 & 0.47 & 5612 & 108 & 35.4 & 0.5 & 4.1 & 0.54 & 0.04 \\
M4RHB261 & RHB & D & 12.88 & 42.36 & 1.61 & 5915 & 108 & 52.9 & 0.5 & 6.1 & 0.67 & 0.05 \\
\end{longtable}

\end{document}